\documentclass[aps, prx, reprint,groupedaddress, longbibliography, showpacs]{revtex4-1}

\usepackage{graphicx}
\usepackage{amsmath, amssymb, xfrac}
\usepackage{threeparttable}
\usepackage{color}
\usepackage{placeins}

\begin{document}


\title{Cavity-excited Huygens' metasurface antennas: near-unity aperture efficiency from arbitrarily-large apertures}


\author{Ariel Epstein}
\author{Joseph P. S. Wong}
\author{George V. Eleftheriades}
\affiliation{The Edward S. Rogers Department
of Electrical and Computer Engineering, University of Toronto, Toronto, ON, Canada M5S 2E4
(email: ariel.epstein@utoronto.ca; gelefth@waves.utoronto.ca).}


\date{\today}

\begin{abstract}
One of the long-standing problems in antenna engineering is the realization of highly-directive beams using low-profile devices. In this paper we provide a solution to this problem by means of Huygens' metasurfaces (HMSs), based on the equivalence principle. This principle states that a given excitation can be transformed to a desirable aperture field by inducing suitable electric and magnetic surface currents. Building on this concept, we propose and demonstrate cavity-excited HMS antennas, where the \textcolor{black}{single-source cavity excitation} is designed to optimize aperture illumination, while the HMS facilitates the current distribution that ensures phase purity of aperture fields. The HMS breaks the coupling between the excitation and radiation spectrum typical to standard partially-reflecting surfaces, allowing tailoring of the aperture properties to produce a desirable radiation pattern.
\textcolor{black}{As shown, a single semianalytical formalism can be followed to achieve control of a variety of radiation features, such as the direction of the main beam or the side lobe level, by proper modification of the HMS and the source position. Relying on a cavity excitation, this can be achieved without incurring edge-taper losses and without any degradation of the aperture illumination for arbitrarily-large apertures. With the recent demonstrations of Huygens' metasurfaces at microwave, terahertz, and optical frequencies, the proposed low-profile design may find its use in a myriad of applications across the electromagnetic spectrum, from highly-directive antennas to highly-efficient quantum-dot emitters, reaching near-unity aperture efficiencies.}

\end{abstract}


\maketitle

\section{Introduction}
\label{sec:introduction}
Achieving high directivity with compact radiators has been a major concern of the antenna community since its early days \cite{Kraus1940,Trentini1956,Hansen1961}. Still today, many modern applications, such as automotive radars, satellite communication, millimetre-wave point-to-point communication, and microwave imaging, strive for simple and efficient low-profile antennas producing the narrowest possible beams \cite{Franson2009,Tichit2011,Jiang2011,Lier2011}. Extending the size of the radiating aperture leads to an enhanced directivity, but only if the aperture is efficiently excited. To date, uniform illumination of large apertures is achievable with reflectors and lenses; although these can be made compact using concurrent metamaterial concepts, they still require substantial separation between the source and the aperture, resulting in a large overall antenna size \cite{Balanis2008_Chap5_Reflectors,Hum2014}. In addition, feed blockage and spillover effects must also be considered, usually complicating the design and reducing the device efficiency. High aperture efficiencies can also be achieved using antenna arrays \cite{Balanis2008_Chap11_Arrays}; nevertheless, the requirement for elaborated feed network significantly increases the complexity of this solution and limits its compactness \cite{Haupt2015}, and may also introduce considerable feed-network losses.

Leaky-wave antennas (LWAs), on the other hand, can produce directive beams using a low-profile structure fed by a simple single source \cite{Balanis2008_Chap7_LWAs}. Their typical configuration consists of a guiding structure with a small perturbation, facilitating coupling of guided modes to free-space radiation. In the much-discussed Fabry-P\'{e}rot (FP) LWAs, a localized source is sandwiched between a perfect electric conductor (PEC) and a partially-reflecting surface (PRS), forming a longitudinal FP cavity \cite{Trentini1956,Jackson2011}. By tuning the cavity height at the design frequency $\omega$, favourable coupling of the source to a single parallel-plate waveguide mode is achieved, forming a dominant leaky wave emanating from the source; the typical device thickness lies around half of a wavelength. The leaky mode is characterized by a transverse wavenumber whose real part $k_t$ corresponds to the waveguide dispersion, and is accompanied by a small imaginary part $\alpha$ determined by the PRS. 
This leads to an azimuthally-symmetric directive radiation through the PRS towards the direction defined by $\sin\theta_\mathrm{out}\approx k_t/k$, where $k=\omega\sqrt{\mu\varepsilon}$ is the free-space wavenumber, with a beamwidth proportional to $\alpha$. Broadside radiation is achieved when $\theta_\mathrm{out}$ is small enough such that the splitting condition $k_t<\alpha$ is satisfied, and the peaks of the conical beam merge \cite{Lovat2006_1}.

Another class of LWAs which has received significant attention lately is based on modulated impedance metasurfaces (MoMetAs) \cite{Fong2010,Minatti2011,Patel2011,Minatti2015}. These so-called holographic antennas use a point source to excite surface waves on a thin dielectric sheet covered with metallic patches and backed by a PEC ground plane, establishing effective surface impedance boundary conditions \cite{Sievenpiper1999}; guiding surface waves, these structures can be very thin, below fifth of a wavelength. Similar to FP-LWAs, the guiding structure is designed such that only a single surface mode $k_t>k$ is allowed to propagate; small modulation of the surface impedance, implemented by variation of patch sizes or dielectric thickness, couples the bound modes to radiative modes. To facilitate such coupling in the case of surface waves, whose transverse momentum is greater than that of free-space, the impedance modulation should have a periodicity $a=2\pi/K$ comparable to the wavelength. The interaction between the surface wave and the perturbation results in an infinite number of Floquet-Bloch (FB) harmonics $k_{t,n}=k_t+nK$; the periodicity should be designed such that one of them radiates to the desirable direction, while the others become evanescent, ensuring good directivity. The leakage rate $\alpha$, and correspondingly the beamwidth, are determined by the depth of the modulation \cite{Minatti2015}.

Both FP-LWAs and MoMetAs have an appealing compact configuration and their radiation characteristics can be rather simply controlled by tuning the properties of the guiding structure and the perturbation. Nonetheless, due to their leaky-wave nature, they suffer from a fundamental efficiency limitation when considering practical finite apertures: designing a moderate leakage rate $\alpha$ with respect to the aperture length $L$ yields uniform illumination of the aperture (high aperture efficiency) but results in considerable losses from the edges (low radiation efficiency); on the other hand, large values of $\alpha$ lead to high radiation efficiencies but in this case only a portion of the aperture is used for radiation, leading to a wider beam \cite{Sievenpiper2005,Komanduri2010,GarciaVigueras2012}. 

To mitigate edge-taper losses, shielded FP-LWA structures have been recently proposed, using PEC side walls which form a lateral cavity  
\cite{Feresidis2006, Ju2009, Muhammad2012, Muhammad2014, Kim2012_1, Hosseini2013, Haralambiev2014}. Nevertheless, the tight coupling between the propagation of the leaky mode inside the FP cavity and the angular distribution of the radiated power manifested by $\sin\theta_\mathrm{out}\approx k_t/k$ poses serious limitations on the achievable aperture efficiency. This is most prominent for antennas radiating at broadside, in which only low-order lateral modes, carrying transverse wavenumbers which are small enough to satisfy the splitting condition, can be used. Consequently, such antennas are designed to excite exclusively the $\mathrm{TE_{10}}$ lateral mode, which inherently limits the aperture efficiency, defined as the relative directivity with respect to the case of uniform illumination, 
to about $80\%$ \cite{Balanis1997_Chap12}. Although this problem can be partially solved by the use of artificial magnetic conductor (AMC) side walls instead of PECs \cite{Ronciere2007, Kelly2009}, 
one of the most limiting constraints on the antenna design which still remains is the requirement for mode purity. As the dominant spectral components of the field in the cavity directly translate to prominent lobes in the radiation pattern, only a single mode should be allowed to propagate in the cavity to guarantee high directivity. However, as demonstrated by \cite{Muhammad2011}, suppression of parasitic modes in a cavity is usually a very difficult problem. In particular, single mode excitation becomes increasingly challenging as the desirable aperture size increases, due to the small differences between the wavenumbers associated with low-order modes; thus, in practice, this solution cannot be used for arbitrarily-large apertures.

From the discussion so far it follows that it would be very beneficial if we could optimize separately the fields inside the cavity and the fields formed on the aperture. This would allow us to achieve good illumination of the aperture without the necessity to meet restricting conditions (e.g., the splitting condition, or single mode excitation), stemming from the coupling between excitation and radiation fields. But how to achieve such a separation? The equivalence principle suggests that for a given field exciting a surface, desirable (arbitrary) aperture fields can be formed by inducing suitable electric and magnetic surface currents, supporting the required field discontinuities \cite{Balanis1997_Chap12}. Based on this idea, the concept of Huygens' metasurfaces (HMSs) has been recently proposed as a means for versatile wavefront manipulation \cite{Pfeiffer2013,Selvanayagam2013,Selvanayagam2014,Radi2014,Zhu2014,Pfeiffer2014,Wong2014,Wong2015,Decker2015}. 

Huygens' metasurfaces are planar structures composed of subwavelength elements (meta-atoms), engineered to generate the surface currents required by the equivalence principle to achieve a prescribed functionality. In general, for a given incident field and desirable transmitted field, elements exhibiting effective loss and gain are required for the implementation \cite{Selvanayagam2013_1, Kim2014}. However, for certain applications, the fields can be judiciously stipulated such that the metasurface can be constructed from passive and lossless elements, i.e. electric and magnetic polarizable particles \cite{Pfeiffer2013, Selvanayagam2013}. In fact, we have recently shown that if the reflected and transmitted fields are set such that the wave impedance and the real power are continuous across the two facets of the metasurface, the aperture phase can be tailored by a passive and lossless HMS to produce directive radiation towards a prescribed angle $\theta_\mathrm{out}$, for any given excitation source; the design procedure is straightforward once the source plane-wave spectrum is assessed \cite{Epstein2014}.

\begin{figure}[htb]
\centering
\includegraphics[width=8cm]{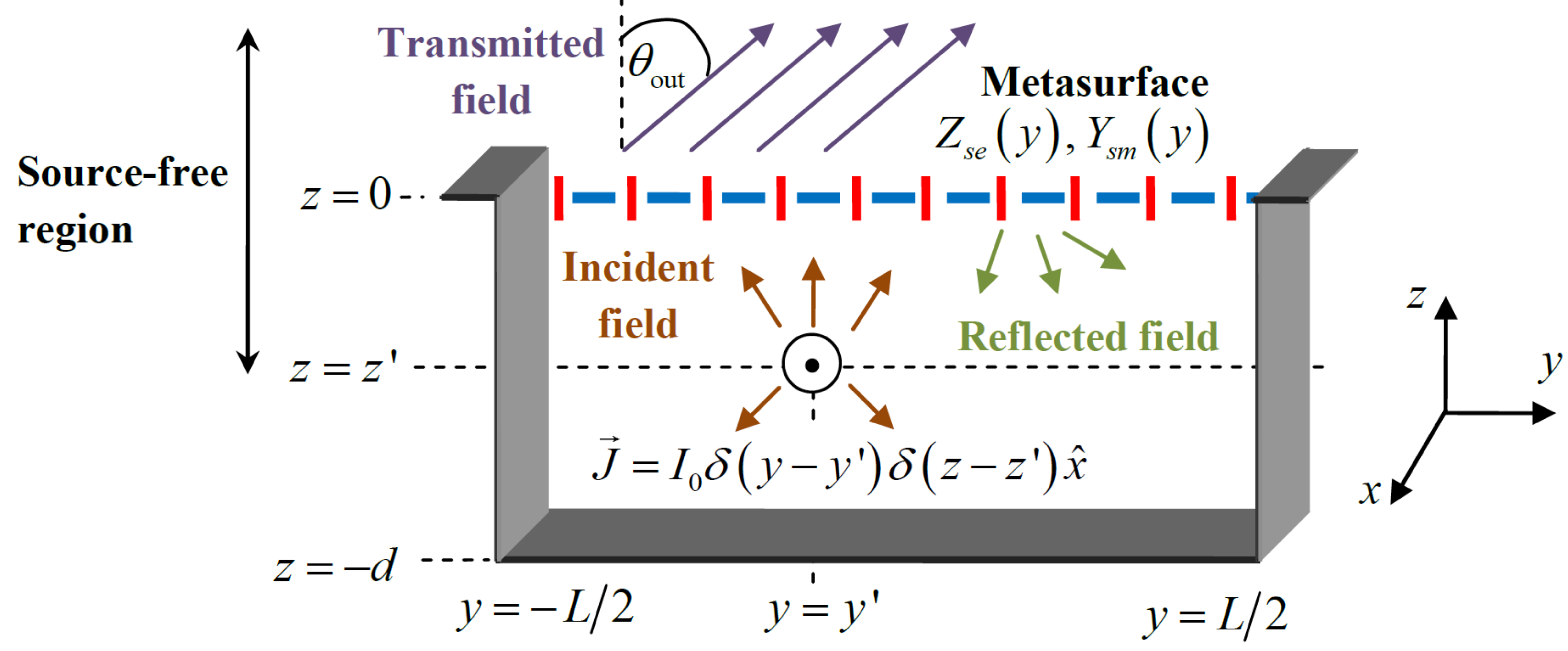}
\caption{{Physical configuration of a cavity-excited Huygens' metasurface antenna.}}
\label{fig:physical_configuration}
\end{figure}

Indeed, in this paper we propose to harness the equivalence principle to efficiently convert fields excited in a cavity by a localized source to highly-directive radiation using a Huygens' metasurface: cavity-excited HMS antenna. The device structure resembles a typical shielded FP-LWA configuration, with an electric line source surrounded by three PEC walls and a Huygens' metasurface replacing the standard PRS (Fig. \ref{fig:physical_configuration}). 
For a given aperture length $L$ and a desirable transmission angle $\theta_\mathrm{out}$, we optimize the FP cavity thickness and source position to predominantly excite the highest-order mode of the lateral cavity, with the HMS reflection coefficient ensuring the wave impedance is equalized along the metasurface; this guarantees the aperture is well illuminated. Once the source configuration is established, we stipulate the aperture fields to follow the power profile of the cavity mode, ensuring the real power is conserved at each point, and impose the suitable linear phase to promote radiation towards $\theta_\mathrm{out}$. With the cavity fields and aperture fields in hand, we invoke the equivalence principle and evaluate the electric surface impedance and magnetic surface admittance required to support the resultant field discontinuity \cite{Kuester2003,Tretyakov2003,Selvanayagam2013,Pfeiffer2013}. Our previous work \cite{Epstein2014} guarantees that these would be purely reactive, hence could be implemented using passive and lossless meta-atoms.


Utilizing the equivalence principle as described results in formation of aperture fields, the magnitude of which follows the power distribution inside the cavity, whereas their phase is independently determined to vary in a plane-wave-like fashion. This has two important implications. First, as the power profile of the highest-order lateral mode creates hot spots of radiating surface currents approximately half a wavelength apart, a uniform virtual phased-array is formed on the HMS aperture; based on array theory \cite{Balanis2008_Chap11_Arrays}, such excitation profile is expected to yield very high directivity with no grating lobes regardless of the scan angle $\theta_\mathrm{out}$. Second, in contrast to LWAs of any type, the antenna directivity does not deteriorate significantly even if other modes are partially excited, as these would merely vary the amplitude of the virtual array elements, without affecting the phase purity.

This semianalytical design procedure can be applied to arbitrarily-large apertures, yielding near-unity aperture efficiencies, in agreement with full-wave simulations; due to the PEC side walls, no power is lost via the edges. This offers an effective way to overcome the efficiency tradeoff inherent to FP-LWAs and MoMetAs, while preserving the advantages of a single-feed low-profile antenna.

\section{Theory}
\label{sec:theory}
\subsection{Formulation}
\label{subsec:formulation}
To design the HMS-based antenna, we simply apply the general methodology developed in \cite{Epstein2014} to the source configuration of Fig. \ref{fig:physical_configuration}; for completeness, we recall briefly its main steps. We consider a 2D  scenario ($\partial/\partial x=0$) with the HMS at $z=0$ and a given excitation geometry at $z\leq z'<0$ embedded in a homogeneous medium ($k=\omega\sqrt{\epsilon\mu}$, $\eta=\sqrt{\mu/\varepsilon}$). 
Under these circumstances, the incident, reflected and transmitted fields in the vicinity of the HMS can be expressed via their plane-wave spectrum \cite{FelsenMarcuvitz1973}
\begin{align}
&\left\lbrace
\begin{array}{l}
\vspace{5pt}
\!\!\!E_x^{\mathrm{inc}}\left(y,z\right) = k\eta I_0 \mathcal{F}^{-1}\left\lbrace \frac{1}{2\beta}f\left(k_t\right)e^{-j\beta z}\right\rbrace \\
	\vspace{3pt}
\!\!\!E_x^{\mathrm{ref}}\left(y,z\right) = -k\eta I_0 \mathcal{F}^{-1}\left\lbrace \frac{1}{2\beta}\Gamma\left(k_t\right)f\left(k_t\right)e^{j\beta z}\right\rbrace \\
		\vspace{3pt}
\!\!\!E_x^{\mathrm{trans}}\left(y,z\right) = k\eta I_0 \mathcal{F}^{-1}\left\lbrace \frac{1}{2\beta}\overline{T}\left(k_t\right)e^{-j\beta z}\right\rbrace,
\end{array}\!\!\!\!\!
\right. 
\label{equ:transverse_fields_spectral_domain}
\end{align}
where $\mathcal{F}^{-1}\left\lbrace g\left(k_t;z\right)\right\rbrace\triangleq\frac{1}{2\pi}\int_{-\infty}^{\infty}dk_tg\left(k_t;z\right)e^{jk_ty}$ is the inverse spatial Fourier transform of $g\left(k_t;z\right)$, $f\left(k_t\right)$ is the source spectrum, $\Gamma\left(k_t\right)$ is the HMS reflection coefficient, and $\overline{T}\left(k_t\right)\triangleq T\left(k_t\right)\left[1+\Gamma\left(k_t\right)\right]$ is the transmission spectrum. As before, $k_t$ denotes the transverse wavenumber, and the longitudinal wavenumber is $\beta=\sqrt{k^2-k_t^2}$. 
For simplicity, we only consider here transverse electric (TE) fields ($E_z=E_y=H_x=0$); the nonvanishing magnetic field components $H_y, H_z$ can be calculated from $E_x$ via Maxwell's equations.

For a given source spectrum, it is required to determine the reflected and transmitted fields, through the respective degrees of freedom $\Gamma\left(k_t\right)$ and $T\left(k_t\right)$, that would implement the desirable functionality. 
Once the tangential fields on the two facets of the HMS are set, the equivalence principle is invoked to evaluate the required electric and magnetic surface currents to induce them \cite{Balanis1997_Chap12}. The polarizable particles comprising the HMS are then designed such that the average fields acting on them induce these surface currents \cite{Kuester2003, Tretyakov2003}. Analogously, the HMS can be characterized by its electric surface impedance $Z_{se}\left(y\right)$ and magnetic surface admittance $Y_{sm}\left(y\right)$, relating the field discontinuity and the average excitation via the generalized sheet transition conditions (GSTCs) \cite{Pfeiffer2013,Selvanayagam2013,Epstein2014,Kuester2003}, 
\textcolor{black}{
\begin{equation}
\left\lbrace
\begin{array}{l}
\vspace{5pt}
Z_{se}\!\left(y\right)=\dfrac{1}{2}\!\dfrac{\left.E_x\left(\vec{r}\right)\right|_{z\rightarrow0^+}\!\!+\!\left.E_x\left(\vec{r}\right)\right|_{z\rightarrow0^-}}{\left.H_y\left(\vec{r}\right)\right|_{z\rightarrow0^+}\!\!-\!\left.H_y\left(\vec{r}\right)\right|_{z\rightarrow0^-}}
 \\
	\vspace{3pt}
Y_{sm}\!\left(y\right)=\dfrac{1}{2}\!\dfrac{\left.H_y\left(\vec{r}\right)\right|_{z\rightarrow0^+}\!\!+\!\left.H_y\left(\vec{r}\right)\right|_{z\rightarrow0^-}}{\left.E_x\left(\vec{r}\right)\right|_{z\rightarrow0^+}\!\!-\!\left.E_x\left(\vec{r}\right)\right|_{z\rightarrow0^-}}.
\end{array}
\right. 
\label{equ:GSTC}
\end{equation}}

To promote directive radiation towards $\theta_\mathrm{out}$ we require that the aperture (transmitted) fields approximately follow the suitable plane-wave-like relation
\begin{align}
\left.E_x\left(\vec{r}\right)\right|_{z\rightarrow0^+}&\approx \left.Z_{\mathrm{out}}H_y\left(\vec{r}\right)\right|_{z\rightarrow0^+} \nonumber \\
&\approx k\eta I_0 \mathcal{F}^{-1}\left\lbrace \textstyle\frac{1}{2\beta}T\left(k_t\right)\right\rbrace \nonumber \\ &\triangleq k\eta I_0W_0\left(y\right)e^{-jky\sin\theta_\mathrm{out}},
\label{equ:envelope_function}
\end{align}
where $W_0\left(y\right)$ is the aperture window (envelope) function (yet to be determined) and ${Z}_\mathrm{out}=1/Y_\mathrm{out}=\eta/\cos\theta_\mathrm{out}$ is the TE wave impedance of a plane-wave directed towards $\theta_\mathrm{out}$. 

In previous work, we have shown that if the wave impedance and the real power are continuous across the metasurface, these aperture fields can be supported by a passive lossless HMS (purely reactive $Z_{se}$ and $Y_{sm}$) \cite{Epstein2014}. The first condition, local impedance equalization, means that the total (incident+reflected) fields on the bottom facet of the metasurface should exhibit the same wave impedance, i.e. $\left.E_x\left(\vec{r}\right)\right|_{z\rightarrow0^-}= \left.Z_{\mathrm{out}}H_y\left(\vec{r}\right)\right|_{z\rightarrow0^-}$; this is achieved by setting the reflection coefficient to a Fresnel-like form
\begin{equation}
\Gamma\left(k_t\right)=\frac{k\cos\theta_\mathrm{out}-\beta}{k\cos\theta_\mathrm{out}+\beta}, 
\label{equ:reflection_coefficient}
\end{equation}
determining the reflected fields everywhere \textcolor{black}{[Eq. \eqref{equ:transverse_fields_spectral_domain}]}, fixing our first degree of freedom.
To satisfy the second condition, local power conservation, we require that the aperture window function follows 
the \emph{magnitude} of the total (incident+reflected) fields at $z\rightarrow0^-$, namely, 
\begin{align}
\textstyle W_0\left(y\right) &=\left|E_x\left(\vec{r}\right)\right|_{z\rightarrow0^-}= \left|\mathcal{F}^{-1}\left\lbrace \textstyle \frac{1}{2\beta}\left[1-\Gamma\left(k_t\right)\right]f\left(k_t\right)\right\rbrace\right| \nonumber \\
&= \left(\mathcal{F}^{-1}\left\lbrace \textstyle \left[\frac{1}{2\beta}\left(1-\Gamma\right)f\right]\star\left[\frac{1}{2\beta}\left(1-\Gamma\right)f\right]\right\rbrace\right)^{1/2}
, 
\label{equ:aperture_window_function}
\end{align}
where $g\star g$ is the autocorrelation of the spectral-domain function $g\left(k_t\right)$ \cite{Howell2000}; this determines the transmitted fields everywhere \textcolor{black}{[Eq. \eqref{equ:transverse_fields_spectral_domain}]}, fixing our second degree of freedom. 

The absolute value operator in the last equality is of utmost significance: it indicates that the transmission spectrum of the aperture fields follows, up to a square root, the \emph{power spectral density} (PSD) of $\left.E_x\left(\vec{r}\right)\right|_{z\rightarrow0^-}$, and \emph{not} the spectral content of the incident and reflected \emph{fields}. This is directly related to the balanced (plane-wave-like) contribution of the electric and magnetic fields to the power flow that we stipulated in Eq. \eqref{equ:envelope_function}, and results in a significantly favourable plane-wave spectrum, as will be discussed in detail in the next subsection. 

Finally, we use these semianalytically predicted fields \textcolor{black}{[Eq. \eqref{equ:transverse_fields_spectral_domain}]} and the equivalence principle, manifested by the GSTCs \textcolor{black}{[Eq. \eqref{equ:GSTC}]}, to calculate the required HMS surface impedance, yielding the desirable purely reactive modulation given by,
\begin{equation}
\frac{Z_{se}\left(y\right)}{Z_\mathrm{out}}=\frac{Y_{sm}\left(y\right)}{Y_\mathrm{out}}=-\frac{j}{2}\cot\left[\frac{\phi_{-}\left(y\right)-\phi_{+}\left(y\right)}{2}\right] 
\label{equ:HMS_design}
\end{equation} 
where $\phi_{\pm}\left(y\right)\triangleq \angle\left.E_x\left(y,z\right)\right|_{z\rightarrow 0^{\pm}}$ are the phases of the stipulated fields just above and below the metasurface \cite{Epstein2014}.

Once the general design procedure is established, applying it to the configuration of Fig. \ref{fig:physical_configuration}, which includes an electric line source at $\left(y',z'\right)$ surrounded by PEC walls at $z=-d$, $y=\pm L/2$, is straightforward: it reduces to finding the corresponding source spectrum. The latter is quantized due to the lateral cavity, and includes multiple reflections between the HMS at $z=0$ and the PEC at $z=-d$ \cite{FelsenMarcuvitz1973}; explicitly, 
\begin{equation}
f\left(k_t\right)\!=\!\frac{\pi}{2L}\!\!\! \sum\limits_{n=-\infty}^{\infty} \!\!\!\left\lbrace \!\!\!\!
\begin{array}{l}
\dfrac{e^{-j\beta\left(d+z'\right)}-e^{j\beta\left(d+z'\right)}}{e^{j\beta d}-\Gamma\left(k_t\right)e^{-j\beta d}}\\ 
\left[e^{-jk_ty'}+\left(-1\right)^{n+1}e^{jk_ty'}\right]
\delta\left(k_t-\frac{n\pi}{L}\right) \!\!\!\!
\end{array}
\right\rbrace. 
\label{equ:cavity_source_spectrum}
\end{equation}
We refer to the sum of the fields corresponding to the $n,-n$ terms in the summation as the field of the $n$th mode of the lateral cavity, where $n\geq0$.

Although this procedure is applicable for any transmission angle, we restrict ourselves from now on to the case of broadside radiation $\theta_\mathrm{out}=0$, where the performance of shielded and unshielded FP-LWAs is the most problematic due to the splitting condition \cite{Lovat2006_1} (design of oblique-angle radiators is addressed in Appendix \ref{sec:oblique_angle}). In this case, the desirable radiation pattern is symmetric with respect to the $\widehat{xz}$ plane, thus it is only natural to use a laterally-symmetric excitation; subsequently, we set the lateral position of the source to be $y'=0$ throughout the rest of the paper. With this selection, the even modes vanish, and the odd modes follow a cosine profile in the lateral dimension. 

\subsection{Optimizing the cavity excitation}
\label{subsec:cavity_design}
One of the key differences between the cavity-excited HMS antenna and FP-LWAs is that by harnessing the equivalence principle we control the \emph{individual} contributions of the electric and magnetic fields to the flow of power, expressed by the lateral distribution of the $z$-component of the Poynting vector on the aperture. More specifically, as mentioned in the previous subsection, the resultant (transmitted) aperture fields corresponding to Eq. \eqref{equ:aperture_window_function} actually follow the square root of the \emph{power} profile dictated by the cavity mode, and \emph{not} the profile of the cavity \emph{fields}. This distinction is very important, as the \emph{power} profile of a standing wave is always positive, whereas the \emph{field} profile changes signs along the lateral dimensions. Hence, the spectral content of the aperture fields, which determines the far-field radiation pattern, is fundamentally different.

\begin{figure*}[htb]
\centering
\includegraphics[width=17cm]{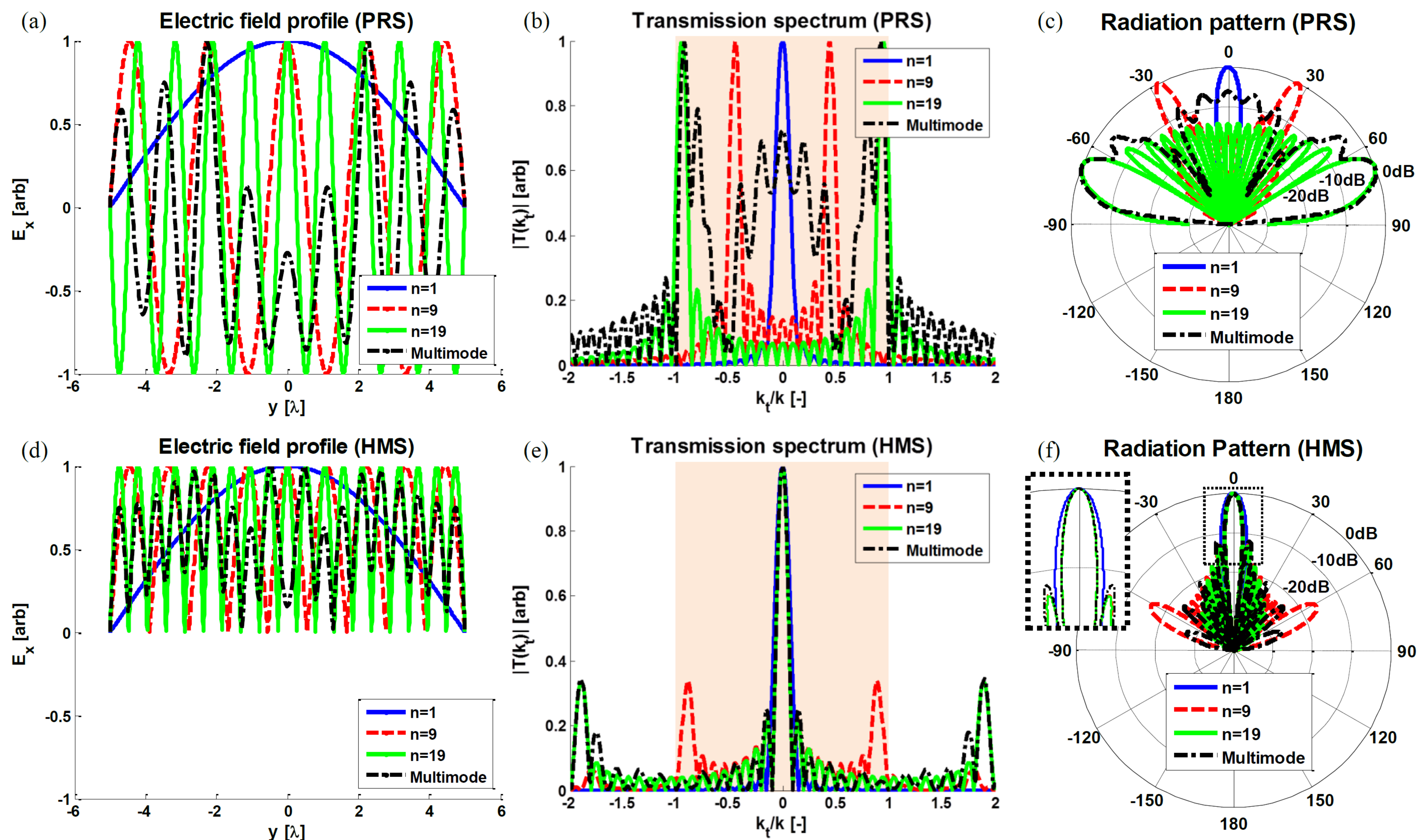}
\caption{{Comparison between aperture profiles of cavity-excited PRS, as in shielded FP-LWAs, and cavity-excited HMS, and their implication on the radiation patterns.} \emph{Single mode} excitations of the $n=1$ (blue solid line), $n=9$ (red dashed line), and $n=19$ (green solid line) modes of an aperture of length $L=10\lambda$ are compared to the \emph{multimode} excitation corresponding to the HMS antenna presented in Fig. \ref{fig:physical_configuration} with $L=10\lambda$, $z'=-\lambda$, and $d=1.61\lambda$ (black dash-dotted line). ({a}),({d}) Normalized spatial profile of the tangential electric field on the aperture. ({b}),({e}) Normalized spectral content of the aperture field; shaded region correspond to the visible part of the spectrum. ({c}),({f}) Normalized radiation patterns. {Inset:} Close-up of the radiation pattern around $\theta=0$.}
\label{fig:mode_analysis_PRS_HMS}
\end{figure*}

To illustrate this point, and to elucidate the guidelines for optimizing the cavity excitation, we compare the fields formed on the device aperture for a shielded FP-LWA, where a standard PRS is used, and for a cavity-excited HMS antenna with the same excitation. Figure \ref{fig:mode_analysis_PRS_HMS} presents the spatial profile of the tangential electric field, its spatial Fourier transform, and corresponding radiation patterns (calculated following \cite{Balanis1997_Chap12}), for single mode excitation of the $n=1$ (solid blue lines), $n=9$ (dashed red lines), and $n=19$ (green solid lines) modes, for an aperture length of $L=10\lambda$. All plots are normalized to their maximum, as the radiation pattern is sensitive to the variation of the fields, and not to their magnitude. 

As follows from Eq. \eqref{equ:cavity_source_spectrum}, the spatial profile of the $n$th-mode aperture field is proportional to $\cos\left(n\pi y/L\right)$ for a standard PRS, but for an HMS designed according to Eq. \eqref{equ:aperture_window_function}, it is proportional to $\left|\cos\left(n\pi y/L\right)\right|$ [Fig. \ref{fig:mode_analysis_PRS_HMS}(a),(d)]. Except for the lowest-order mode $n=1$, for which the two functions coincide, the difference in the spatial profile translates into distinctively different features in the spectral content [Fig. \ref{fig:mode_analysis_PRS_HMS}(b),(e)]. For the $n$th mode, the transmission spectrum of the HMS aperture corresponds to the autocorrelation of the PRS aperture spectrum, leading to formation of peaks centered around the second harmonics ($k_t=\pm2n\pi/L$) and DC ($k_t=0$). As both the right-propagating and left-propagating components of the standing wave coherently contribute to the DC peak, the latter dominates the transmission spectrum, and the radiation patterns corresponding to the HMS aperture exhibit highly-directive radiation towards broadside [Fig. \ref{fig:mode_analysis_PRS_HMS}(f)]. In contrast, the PRS aperture devices exhibit symmetric (conical) radiation to angles determined by the dominant spectral components of the aperture fields \cite{Balanis2008_Chap7_LWAs,Jackson2011}, i.e. towards $\theta=\pm\arcsin\left[n\lambda/\left(2L\right)\right]$ [Fig. \ref{fig:mode_analysis_PRS_HMS}(c)].

The transverse wavenumber $k_t=\pi/L$ corresponding to the lowest-order mode $n=1$ is small enough such that 
the two symmetric beams merge \cite{Lovat2006_1}, which enables the PRS aperture to radiate a single beam at broadside. Indeed, small-aperture shielded FP-LWAs utilize this $\mathrm{TE}_{10}$ mode to generate broadside radiation. However, as demonstrated by \cite{Balanis1997_Chap12}, the aperture efficiency of this mode is inherently limited to about $80\%$, due to the non-optimal cosine-shaped aperture illumination of the lowest-order mode \cite{Feresidis2006, Ju2009, Muhammad2012, Muhammad2014, Kim2012_1, Hosseini2013, Haralambiev2014}, leading to broadening of the main beam [inset of Fig. \ref{fig:mode_analysis_PRS_HMS}(f)]. This highlights a key benefit of using an HMS-based antenna, as it is clear from Fig. \ref{fig:mode_analysis_PRS_HMS}(f) that we can use high-order mode excitations, which provide a more uniform illumination of the aperture, for generating narrow broadside beams with enhanced directivities.

In fact, as the index $n$ of the mode exciting the HMS increases, the autocorrelation of Eq. \eqref{equ:aperture_window_function} drives the second harmonic peaks outside the visible region of the spectrum [shaded region in Fig. \ref{fig:mode_analysis_PRS_HMS}(b),(e)], funnelling all the radiated power to the broadside beam, subsequently increasing the overall directivity. This improvement in radiation properties can be explained using ordinary array theory. As seen from Fig. \ref{fig:mode_analysis_PRS_HMS}(d), the peaks of the field profile generated by the $n$th mode on the HMS aperture form hot spots of radiating currents separated by a distance of $L/n$. The radiation from such an aperture profile would resemble the one of a uniform array with the same element separation. As known from established array theory, to avoid grating lobes the element separation should be smaller than a wavelength \cite{Balanis2008_Chap11_Arrays}. For an aperture length of $L=N\lambda$, where $N$ is an integer, the hot spot separation satisfies this condition for mode indices $n>N$; 
specifically, for $N=10$ (Fig. \ref{fig:mode_analysis_PRS_HMS}), grating lobes would not be present in the radiation pattern for mode indices $n>10$. In agreement with this argument, Fig. \ref{fig:mode_analysis_PRS_HMS}(f) shows that for $n=9$ grating lobes still exist, while for the highest-order mode $n=19$ they indeed vanish.

\begin{figure}[htb]
\centering
\includegraphics[width=7cm]{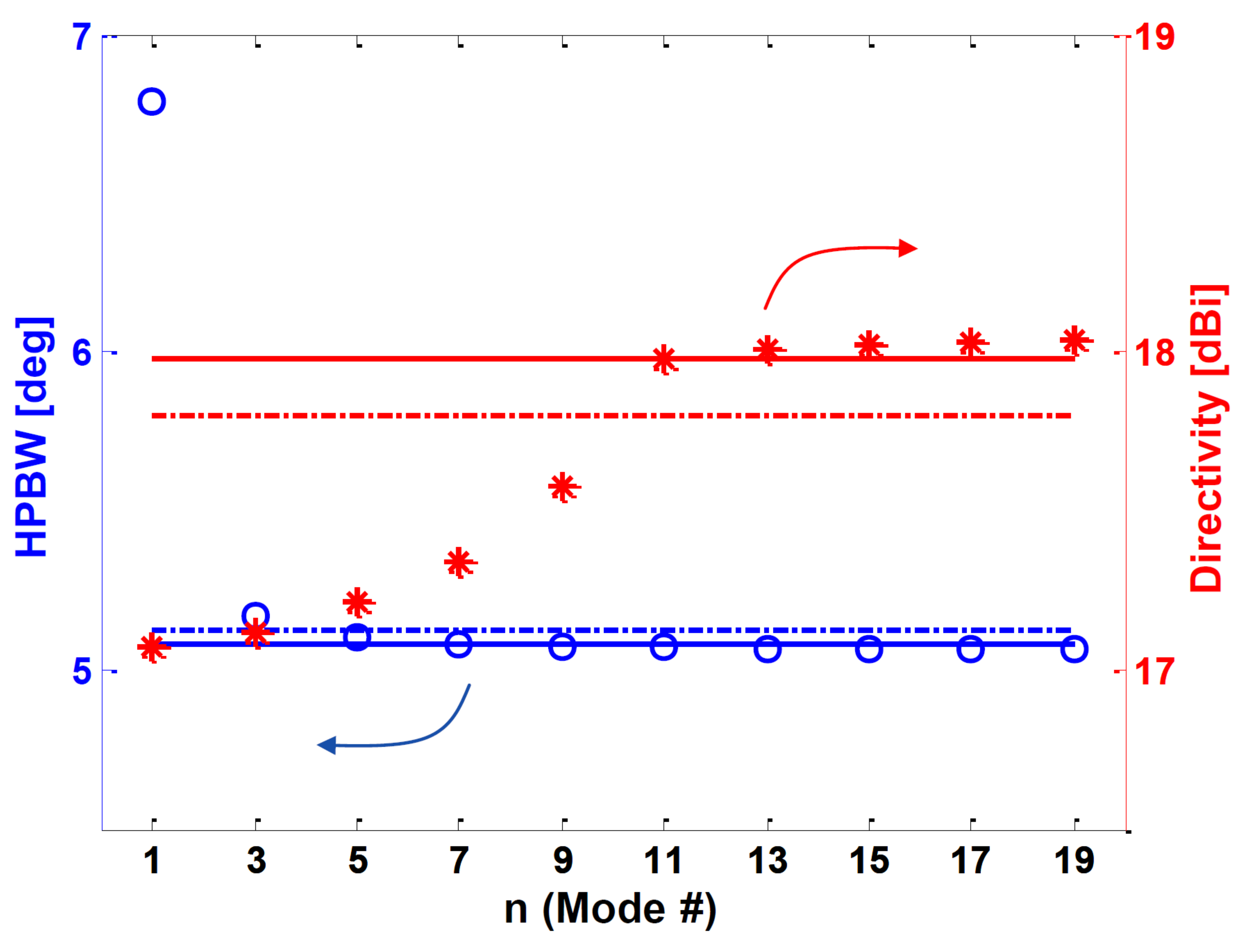}
\caption{{Radiation characteristics of different lateral cavity modes.} Half-power beamwidth (blue circles) and 2D directivity \cite{Lovat2006} (red asterisks) of an HMS aperture of length $L=10\lambda$ excited by a \emph{single} mode as a function of the mode index $n$. Solid lines denote the respective radiation characteristics of a uniformly excited aperture \cite{Balanis1997_Chap12} and dash-dotted lines mark the HPBW (blue) and directivity (red) of \emph{multimode} excitation corresponding to the HMS antenna of Fig. \ref{fig:physical_configuration} with $L=10\lambda$, $z'=-\lambda$, and $d=1.61\lambda$.}
\label{fig:mode_performance}
\end{figure} 

These observations are summarized in Fig. \ref{fig:mode_performance}, where the radiation characteristics of an HMS aperture of $L=10\lambda$ excited by a single mode are plotted as a function of the mode index $n$ (only fast modes $k_{t,n}<k$ are considered). For comparison, solid lines denoting the half-power beamwidth (HPBW) and directivity values achieved for a uniformly-illuminated aperture of the same size are presented as well. Indeed, it can be seen that the lowest-order lateral mode exhibits the worst performance by far, and the performance improves as the mode index increases. While the half-power beamwidth saturates quickly, the directivity values continue to increase with $n$ until the point in which grating lobes disappear $n=N=10$ is crossed; for mode indices $n>10$ the radiation characteristics of the HMS aperture are comparable with those of the optimal uniformly-excited aperture. 

From an array theory point of view, excitation of the highest-order mode is preferable, as the corresponding equivalent element separation approaches $\lambda/2$, implying that such aperture profile would be suitable for directing the radiation to large oblique angles $\theta_\mathrm{out}\neq 0$ without generating grating lobes \cite{Balanis2008_Chap11_Arrays}. Another reason to prefer excitation of the highest-order mode in the case of cavity-excited HMS antennas is that the HMS reflection coefficient $\Gamma\left(k_t\right)$ grows larger with $k_t=n\pi/L$ \textcolor{black}{[Eq. \eqref{equ:reflection_coefficient}]}; therefore, the power carried by the highest-order mode $n=2N-1$ is best-trapped in the cavity, guaranteeing uniform illumination even in the case of very large apertures.

Nevertheless, generating a single-mode excitation of a cavity via a localized source can be very problematic \cite{Chung1971,Muhammad2011}. Fortunately, the cavity-excited HMS antenna can function very well also with multimode excitation, as long as high-order modes dominate the transmission spectrum. This is demonstrated by the dot-dashed lines in Figs. \ref{fig:mode_analysis_PRS_HMS} and \ref{fig:mode_performance}, corresponding to a multimode excitation generated by the configuration depicted in Fig. \ref{fig:physical_configuration} with $L=10\lambda$, $z'=-\lambda$, and $d=1.61\lambda$. 

As expected from the expression for the source spectrum [Eq. \eqref{equ:cavity_source_spectrum}], for a given aperture length $L$, the field just below the aperture due to a line source would be a superposition of lateral modes, the weights of which are determined by the particular source configuration, namely the cavity thickness $d$ and source position $z'$. The multimode transmission spectrum in Fig. \ref{fig:mode_analysis_PRS_HMS}(b) indicates that for the chosen parameter values, high-order modes ($k_t\rightarrow \pm k$) predominantly populate the aperture spectrum, however low-order modes ($k_t\rightarrow 0$) are present as well, to a non-negligible extent (this takes into account the fact that the transmission coefficient ${1-\Gamma\left(k_t\right)}$ is higher for lower-order modes). Considering that the far-field angular power distribution $S\left(\theta\right)$ is proportional to $\cos^2\theta\left|T\left(k_t=k\sin\theta\right)\right|^2$, the multimode excitation of the PRS aperture results in a radiation pattern resembling the one corresponding to single mode excitation of the highest-order mode ($n=19$) but with significant lobes around broadside [Fig. \ref{fig:mode_analysis_PRS_HMS}(c)]; consequently, even if a conical beam is desirable, multimode excitation would result in significant deterioration of the directivity.

On the other hand, the same multimode excitation does not degrade substantially the performance of the HMS antenna. The autocorrelated spectrum, relevant to the field induced on the HMS aperture, results in merging of all spectral components into a sharp DC peak, with the high-order grating lobes pushed to the evanescent region of the spectrum [Fig. \ref{fig:mode_analysis_PRS_HMS}(e)]. This retains a beamwidth comparable with that resulting from a single-mode excitation of the highest-order mode, with only slight deterioration of the directivity due to increased side-lobe level [Fig. \ref{fig:mode_analysis_PRS_HMS}(f) and inset]. Continuing the analogy to array theory, such multimode excitation introduces slight variations to the magnitude of the array elements, forming an equivalent non-uniform array \cite{Balanis2008_Chap11_Arrays}. The corresponding multimode HPBW and directivity values are marked, respectively, by blue and red dash-dotted lines in Fig. \ref{fig:mode_performance}, verifying that indeed, cavity-excited HMS antennas achieve near-unity aperture efficiencies with a practical multimode excitation; this points out another key advantage of the cavity-excited HMS antenna with respect to shielded FP-LWAs. 

With these observations in hand, we are finally ready to formulate guidelines for optimizing the cavity excitation for maximal directivity. 
For a given aperture length $L=N\lambda$, with respect to Eq. \eqref{equ:cavity_source_spectrum}, we maximize the coupling to the $n=2N-1$ mode (which exhibits the best directivity) by tuning the cavity thickness $d$ as to minimize the denominator of the corresponding coupling coefficient; equally important, we minimize the coupling to the $n=1$ mode (which exhibits the worst directivity) by tuning the source position $z'$ as to minimize the numerator of the corresponding coupling coefficient. To achieve these with minimal device thickness we derive the following design rules
\begin{equation}
d=\frac{\lambda}{2}\frac{2N}{\sqrt{4N-1}}\xrightarrow{N\gg 1}\frac{\lambda}{2}\sqrt{N},\,\,\,\,z'\approx-\left(d-\frac{\lambda}{2}\right).
\label{equ:source_config_design_rules}
\end{equation} 
Although this is somewhat analogous to the typical design rules for (unshielded) FP-LWAs \cite{Jackson2011}, the key difference is that for HMS-based antennas we optimize the source configuration \emph{regardless} of the desirable transmission angle $\theta_\mathrm{out}$. This difference is directly related to the utilization of the equivalence principle for the design of the proposed device, which provides certain decoupling between its excitation and radiation spectrum [\textit{cf.} Fig. \ref{fig:mode_analysis_PRS_HMS}(b),(e)]. This decoupling becomes very apparent when the HMS antenna is designed to radiate towards oblique angles $\theta_\mathrm{out}\neq0$, in which case the same cavity excitation yields optimal directivity as well (See Appendix \ref{sec:oblique_angle}).

Two important comments are relevant when considering these design rules. First, even though following Eq. \eqref{equ:source_config_design_rules} maximizes the coupling coefficient of the highest-order mode and minimizes the coupling coefficient of the lowest-order mode, it does not prohibit coupling to other modes. The particular superposition of lateral modes exhibits a tradeoff between bemawidth and side-lobe level (as for non-uniform arrays \cite{Balanis2008_Chap11_Arrays}). Thus, final optimization of the cavity illumination profile is achieved by fine-tuning the source position $z'$ for the cavity thickness $d$ derived in Eq. \eqref{equ:source_config_design_rules}, with the aid of the efficient semianalytical formulas. In fact, the source position $z'$ is another degree of freedom that can be used to optimize the radiation pattern for other desirable performance features, such as minimal side-lobe level; this feature is further discussed and demonstrated in Appendix \ref{sec:reduced_SLL}. Second, although the optimal device thickness increases with increasing aperture length, the increase is sublinear, following an asymptotic square-root proportion factor. Therefore, applying the proposed concept to very large apertures would still result in a relatively compact device, while efficiently utilizing the aperture for producing highly-directive pencil beams.



\section{Results and Discussion}
\label{sec:results}
We follow the design procedure and the considerations discussed in Section \ref{sec:theory} to design cavity-excited HMS antennas for broadside radiation with different aperture lengths: $L=10\lambda$, $L=14\lambda$, and $L=25\lambda$. The cavity thickness was determined via Eq. \eqref{equ:source_config_design_rules} to be $d=1.61\lambda$, $d=1.89\lambda$, and $d=2.50\lambda$, respectively; the source position was set to $z'=-1.00\lambda$, $z'=-1.33\lambda$, and $z'=-1.94\lambda$, respectively, exhibiting maximal directivity.

\begin{figure}[htb]
\centering
\includegraphics[width=7.5cm]{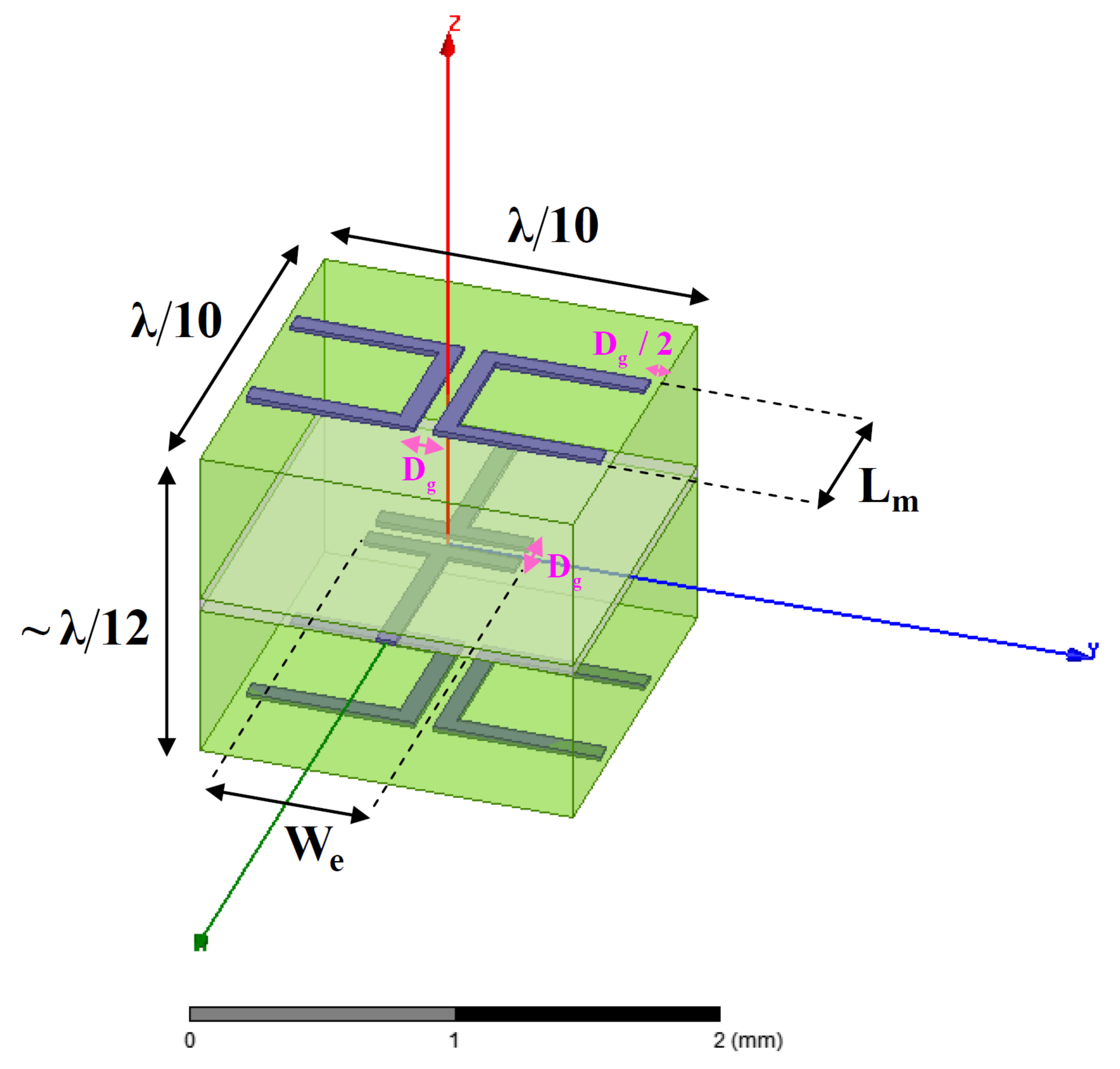}
\caption{Physical configuration of the "spider" unit cells used for implementing the HMS at a frequency of $f=20\mathrm{GHz}$ ($\lambda\approx15\mathrm{mm}$). The electric response is controlled by the capacitor width $W_e$ of the electric dipole, while the magnetic response is determined by the magnetic dipole arm length $L_m$ (See Appendix \ref{sec:spider_cells_modelling} for detailed description).}
\label{fig:unit_cell}
\end{figure} 

The required electric surface impedance and magnetic surface admittance modulations \textcolor{black}{[Eq. \eqref{equ:HMS_design}]} are implemented using the "spider" unit cells depicted in Fig. \ref{fig:unit_cell}. At the design frequency $f=20\mathrm{GHz}$ ($\lambda\approx15\mathrm{mm}$), the unit cell transverse dimensions are $\lambda/10\times\lambda/10$ and the longitudinal thickness is $52\mathrm{mil}\approx\lambda/12$. Each unit cell consists of 3 layers of metal traces defined on two bonded laminates of high-dielectric-constant substrate (See Appendix \ref{sec:spider_cells_modelling}). The two (identical) external layers provide the magnetic response of the unit cell, corresponding to the magnetic surface susceptance $B_{sm}=\Im\left\lbrace Y_{sm}\right\rbrace$, which is tuned by modifying the arm length $L_m$ (affects magnetic currents induced by tangential magnetic fields $H_y$). Analogously, the middle layer is responsible to the electric response of the meta-atom, corresponding to the electric surface reactance $X_{se}=\Im\left\lbrace Z_{se}\right\rbrace$, which is tuned by modifying the capacitor width $W_e$ (affects electric currents induced by tangential electric fields $E_x$). By controlling the lengths of $L_m$ and $W_e$, the spider unit cells can be designed to exhibit Huygens source behaviour, with balanced electric and magnetic responses ranging from $B_{sm}\eta=X_{se}/\eta=-3.1$ to $B_{sm}\eta=X_{se}/\eta=0.9$ (See Appendix \ref{sec:spider_cells_modelling}). 

Figure \ref{fig:antenna_results} presents the design specifications, field distributions, and radiation patterns for the three cavity-excited HMS antennas; Table \ref{tab:antenna_performance} summarizes the antenna performance parameters (for reference, parameters for uniformly-excited apertures \cite{Balanis1997_Chap12} are also included). The semianalytical predictions \textcolor{black}{[Eq. \eqref{equ:transverse_fields_spectral_domain}]} are compared to full-wave simulations conducted with commercially-available finite-element solver (ANSYS HFSS), where the HMS was implemented using the aforementioned spider cells (See Appendix \ref{sec:antenna_simulations}). As demonstrated by Fig. \ref{fig:antenna_results}(a)-(c), the realized unit cells are capable of reproducing the required surface impedance modulation, except maybe around large values of $B_{sm}\eta=X_{se}/\eta$; however, it has been shown that such discrepancies usually have little effect on the performance of Huygens' metasurfaces \cite{Epstein2014_2}. 

\begin{table*}[htb]
\begin{threeparttable}[b]
\renewcommand{\arraystretch}{1.3}
\caption{{Radiation characteristics of cavity-excited HMS antennas} (corresponding to Fig. \ref{fig:antenna_results}).}
\label{tab:antenna_performance}
\centering
\begin{tabular}{l|c c c|c c c|c c c}
\hline \hline
& \multicolumn{3}{c}{$L=10\lambda$ ($d=1.61\lambda, \left|z'\right|=1.00\lambda$)} & \multicolumn{3}{|c}{$L=14\lambda$ ($d=1.89\lambda, \left|z'\right|=1.33\lambda$)} 
	& \multicolumn{3}{|c}{$L=25\lambda$ ($d=2.50\lambda, \left|z'\right|=1.94\lambda$)} \\
\hline
& Full-wave & Semianlytical & Uniform 
& Full-wave & Semianlytical & Uniform 
& Full-wave & Semianlytical & Uniform \\
\hline \hline \\[-1.3em]	
	 \begin{tabular}{l} HPBW \end{tabular}
	 & $5.11^{\circ}$ & $5.13^{\circ}$ & $5.08^{\circ}$
	 & $3.83^{\circ}$ & $3.64^{\circ}$ & $3.63^{\circ}$
	 & $2.09^{\circ}$ & $2.13^{\circ}$ & $2.03^{\circ}$
	 	 	\\	\hline 
	 \begin{tabular}{l} Directivity (2D) [dBi] \end{tabular}
	 & $17.42$ & $17.84$ & $17.98$
	 & $18.79$ & $19.15$ & $19.44$
	 & $21.33$ & $21.75$ & $21.96$
	 	 	 	 \\	\hline 
	 \begin{tabular}{l} First Side-Lobe \end{tabular}
	 & $8.6^{\circ}$ & $8.6^{\circ}$ & $8.2^{\circ}$
	 & $6.4^{\circ}$ & $6.3^{\circ}$ & $5.9^{\circ}$
	 & $3.4^{\circ}$ & $3.4^{\circ}$ & $3.3^{\circ}$
	 	 	 	 \\	\hline 
	 \begin{tabular}{l} Side-Lobe Level [dB] \end{tabular}
	 & $-10.4$ & $-12.0$ & $-13.5$
	 & $-11.1$ & $-10.4$ & $-13.5$
	 & $-14.6$ & $-14.0$ & $-13.5$
	 	 	 	 \\ 	
\hline \hline
\end{tabular}
\end{threeparttable}
\end{table*}

\begin{figure*}[htb]
\centering
\includegraphics[width=14cm]{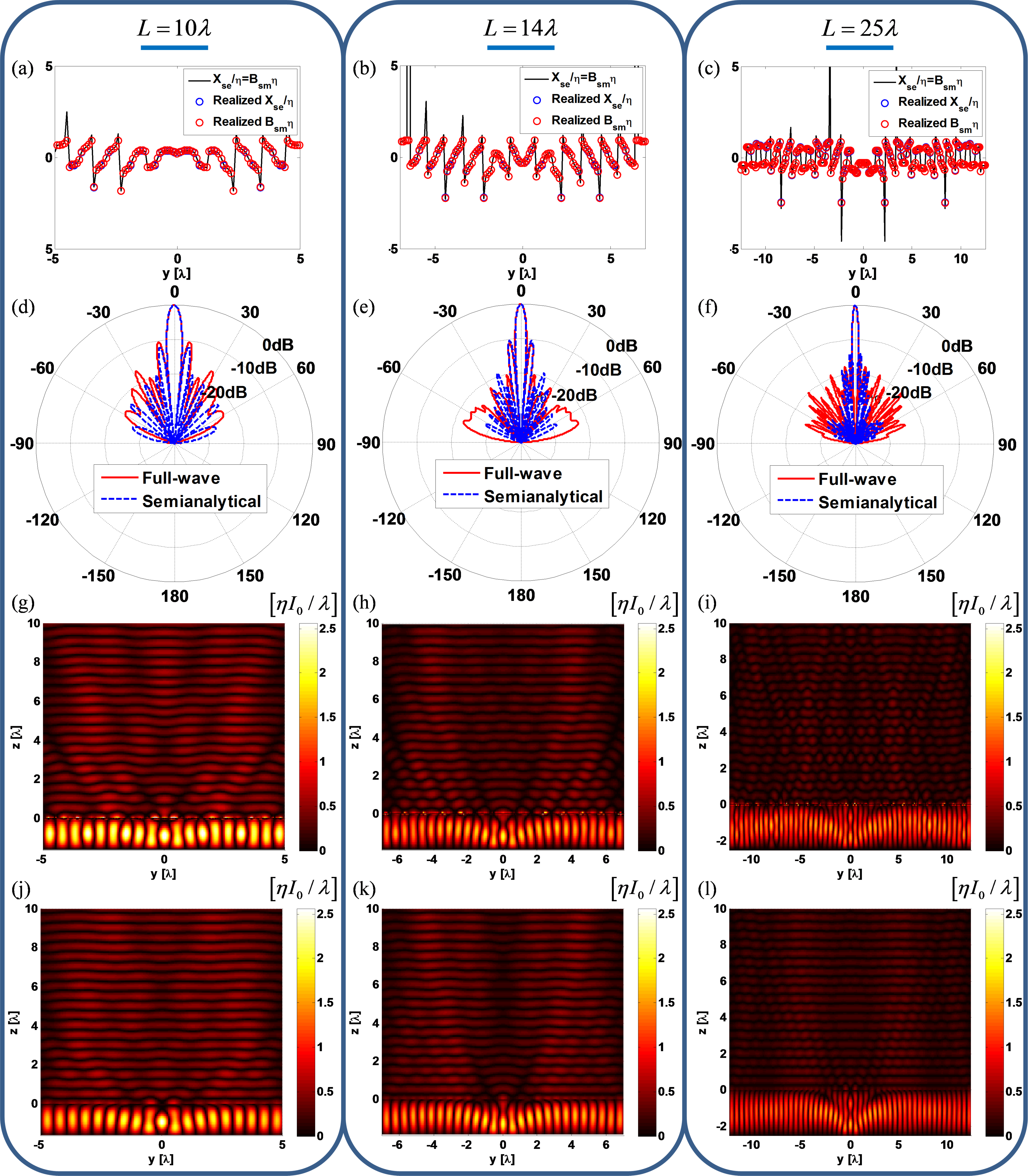}
\caption{Performance of cavity-excited HMS antennas with aperture lengths of $L=10\lambda$, $L=14\lambda$, and $L=25\lambda$. ({a})-({c}) HMS design specifications $X_{se}\left(y\right)/\eta=B_{sm}\left(y\right)\eta$ (black solid line) derived from Eq. \eqref{equ:HMS_design}, and the realized electric surface reactance (blue circles) and magnetic surface susceptance (red circles) using the spider unit cells. ({d})-({f}) Radiation patterns produced by semianalytical formalism (blue dashed line) and full-wave simulations (red solid line). ({g})-({i}) Field distribution $\left|\Re\left\lbrace E_x\left(y,z\right)\right\rbrace\right|$ produced by full-wave simulations. ({j})-({l}) Semianalytical prediction of $\left|\Re\left\lbrace E_x\left(y,z\right)\right\rbrace\right|$ \cite{Epstein2014}.
}
\label{fig:antenna_results}
\end{figure*}
\clearpage

The results in Fig. \ref{fig:antenna_results} and Table \ref{tab:antenna_performance} indicate that the fields and radiation properties predicted by the semianalyical formalism are in excellent agreement with the full-wave simulations for a wide range of aperture lengths. The utilization of realistic (lossy) models for the conductors and dielectrics in the simulated device, as well as other deviations from the assumptions of the design procedure (Appendix \ref{sec:design_assumptions}),  
result in some discrepancies between the full-wave simulations and predicted performance; however, these mostly affect radiation to large angles [Fig. \ref{fig:antenna_results}(d)-(f)]. While this contributes to a minor quantitative difference in the directivity, the properties of the main beam and the side lobes follow accurately the semianalytical results (Table \ref{tab:antenna_performance}), indicating that the theory can reliably predict the dominant contributions to the radiation pattern, as discussed in reference to Fig. \ref{fig:mode_analysis_PRS_HMS}.   

For all cases considered, the excitation of the highest-order lateral mode is clearly visible [Fig. \ref{fig:antenna_results}(g)-(l)], leading to beamwidth and (2D) directivity values $D$ comparable with the ones achieved by uniform excitation of the aperture (Table \ref{tab:antenna_performance}).  
In particular, the simulated radiation patterns yield aperture efficiencies $\eta_\mathrm{apr}\triangleq D/\left(2\pi L/\lambda\right)$ of $88\%$, $86\%$, and $87\%$ for the $L=10\lambda$, $L=14\lambda$, and $L=25\lambda$ devices, respectively, retained even when the aperture length is very large. In terms of half-power beamwidth, often taken as a measure for effective exploitation of the aperture \cite{Sievenpiper2005}, the device performance is even closer to that of a unifromly-excited aperture, with pencil beam HPBWs reaching $99\%$, $95\%$, and $97\%$ of the optimal beamwidth, for $L=10\lambda$, $L=14\lambda$, and $L=25\lambda$ devices, respectively. It should be stressed that even though optimized $\mathrm{TE}_{10}$ shielded FP-LWAs can reach aperture efficiencies of $80\%$, their HPBWs are limited to about $75\%$ of the optimal beamwidth \cite{Balanis1997_Chap12}; more importantly, their PRS-based design requires single-mode excitation to achieve this performance, thus preventing practical realization of large-aperture devices.

We would like to emphasize that this near-optimal aperture utilization is achieved while using realistic models for the substances comprising the metasurface, as well as geometrical dimensions compatible with practical fabrication techniques. It is also very clear from the results that there is no apparent degradation of the structure performance for increasing aperture length, even for the very large values considered. Therefore, it would be reasonable to conclude that the concept introduced in this paper can be applied to design cavity-excited HMS antennas with arbitrarily-large apertures which will exhibit near-unity aperture efficiency.


\section{Conclusion}
\label{sec:conclusion}
We have introduced a novel design for low-profile single-fed antennas exhibiting beamwidth and directivity values comparable with uniformly-excited apertures. To that end, we harness the equivalence principle to devise a cavity-excited Huygens' metasurface, setting the source configuration, HMS reflection coefficient, and aperture fields such that (1) the highest-order lateral mode is predominantly excited, which guarantees that the aperture is well-illuminated; (2) the aperture fields follow the incident \emph{power} profile and \emph{not} the incident \emph{field} profile, which forms an array-like aperture profile with favourable transmission spectrum; and (3) the power flow and wave impedance are continuous across the metasurface, which ensures the design can be implemented by a passive and lossless HMS. The possibility to control the field discontinuities using the electric and magnetic currents induced on the HMS allows us to optimize separately the cavity excitation and the radiated fields, thus overcoming the fundamental tradeoff existing in FP-LWAs between aperture efficiency and edge-taper losses.

It should be emphasized that the general design procedure formulated and demonstrated herein facilitates further optimization of such devices for various applications. The extensive freedom one has in choosing the source configuration, combined with the efficient semianalytical approach, allows explorations of other excitation sources, e.g. with different orientations and current distributions, to tailor the aperture fields for one's requirements, e.g. minimal side-lobe level or radiation towards an oblique angle (See Appendices \ref{sec:oblique_angle} and \ref{sec:reduced_SLL}); once the source spectrum is characterized, the rest of the procedure is straightforward, and the fields and radiation patterns are readily predicted. \textcolor{black}{In particular, the formalism can be readily applied to design cavities based on dielectric Bragg reflectors instead of metallic mirrors, more suitable for terahertz and optical devices. In view of the recent demonstrations of terahertz and optical (Huygens') metasurfaces  \cite{Decker2015,Pfeiffer2014,Monticone2013,Cheng2014,Yu2014,Lin2014,Campione2015,Aieta2015}, this would allow realization of compact and efficient pencil beam radiators across the electromagnetic spectrum, extending the range of applications even further, e.g. for efficient outcoupling of single-photon emission \cite{Lodahl2015} or terahertz elementary sources \cite{Seletskiy2011}}.

\appendix
\section{Cavity-excited HMS antennas radiating at oblique angle}
\label{sec:oblique_angle}
To design cavity-excited HMS antennas radiating at oblique angle we follow the same procedure outlined in Section \ref{sec:theory} for broadside radiators, with the desirable ${{\theta }_{\text{out}}}\ne 0$ inserted into Eqs. \eqref{equ:envelope_function} and \eqref{equ:reflection_coefficient}; note that now ${{Z}_\mathrm{out}}=1/{{Y}_{\text{out}}}=\eta /\cos {{\theta }_{\text{out}}}\ne \eta $. The same argumentations for optimizing the cavity excitation holds (Subsection \ref{subsec:cavity_design}), where the HMS transmission spectra plotted in Fig. \ref{fig:mode_analysis_PRS_HMS}(e) merely shifted by ${{k}_{t}}=k\sin {{\theta }_{\text{out}}}$. This is essentially identical to the design of antenna arrays radiating at an angle ${{\theta }_{\text{out}}}$, where the magnitude of the array element currents are usually determined independently of ${{\theta }_{\text{out}}}$, and the direction of the main beam is manifested via phase-shifts imposed between adjacent elements \cite{Balanis2008_Chap11_Arrays}. Therefore, the optimal configuration still follows Eq. \eqref{equ:source_config_design_rules}, i.e. we use the same cavity excitation regardless of ${{\theta }_{\text{out}}}$. Once the required electric surface impedance and magnetic surface susceptance are specified, the corresponding spider unit-cell dimensions are retrieved from a lookup table constructed as described in Appendix \ref{sec:spider_cells_modelling}, with ${{Z}_\mathrm{out}}=1/{{Y}_{\text{out}}}=\eta /\cos {{\theta }_{\text{out}}}$.

\begin{figure}[htb]
\centering
\includegraphics[width=8cm]{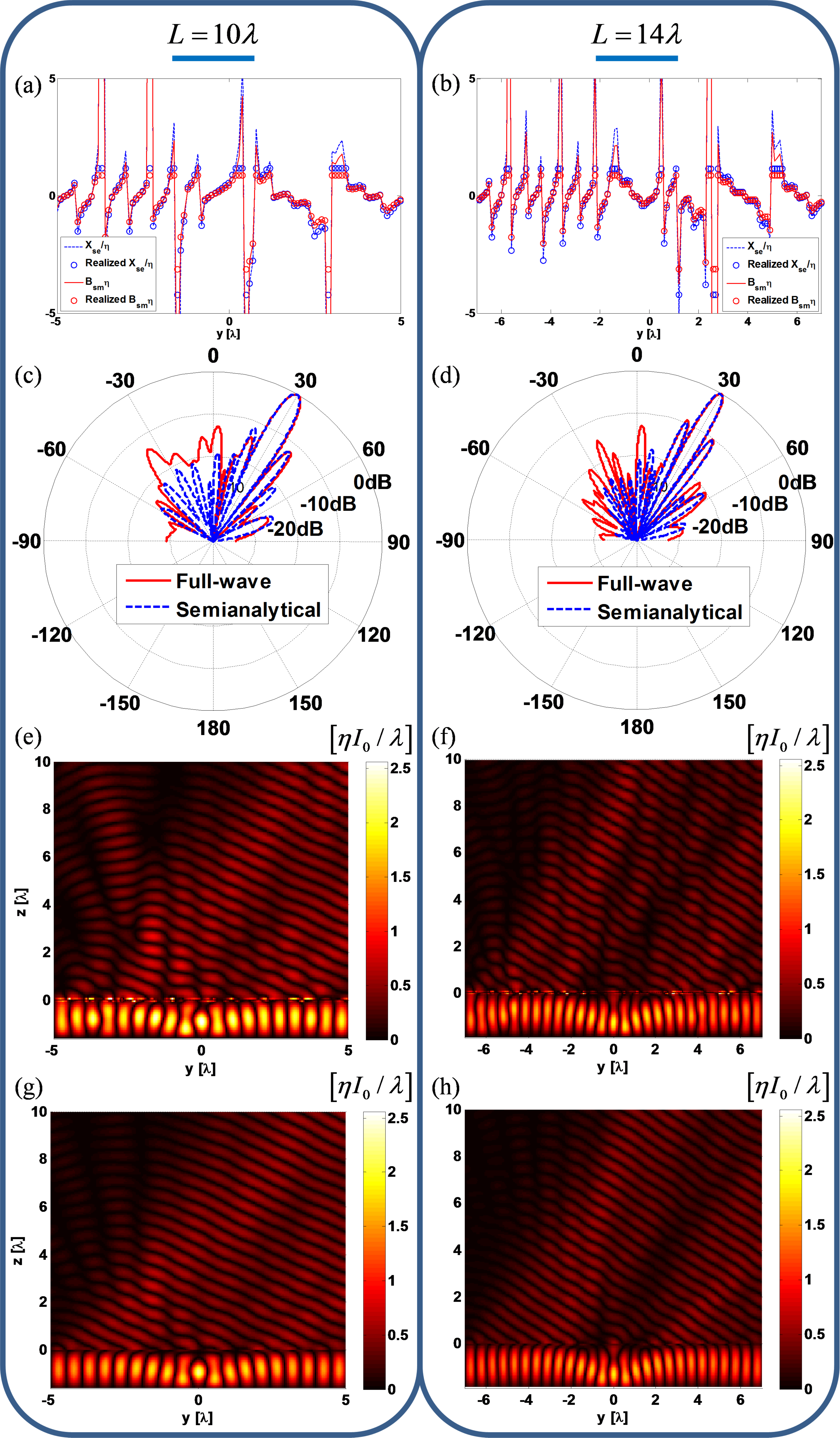}
\caption{Performance of cavity-excited HMS antennas radiating towards $\theta_\mathrm{out}=30^{\circ}$, with aperture lengths of $L=10\lambda$ and $L=14\lambda$. ({a})-({b}) Specified (blue dashed line) and realized (blue open circles) electric surface reactance, and specified (red solid line) and realized (red open circles) magnetic surface susceptance using the spider unit-cells. ({c})-({d}) Radiation patterns produced by semianalytical formalism (blue dashed line) and full-wave simulations (red solid line). ({e})-({f}) Field distribution $\left|\Re\left\lbrace E_x\left(y,z\right)\right\rbrace\right|$ produced by full-wave simulations. ({g})-({h}) Semianalytical prediction of $\left|\Re\left\lbrace E_x\left(y,z\right)\right\rbrace\right|$ \cite{Epstein2014}.
}
\label{fig:antenna_results_oblique}
\end{figure}

We use this outlined procedure to design cavity-excited HMS antennas radiating towards ${{\theta }_{\text{out}}}=30{}^\circ $. 
The design specifications and radiation characteristics are presented in Fig. \ref{fig:antenna_results_oblique} and Table \ref{tab:antenna_performance_oblique} for devices with aperture lengths of $L=10\lambda $ and $L=14\lambda $, comparing the results of full-wave simulations, semianalytical predictions, and uniformly-excited apertures (note that for the latter, HPBW and directivity are factored by $\cos {{\theta }_{\text{out}}}$ with respect to their values for broadside radiation \cite{Sievenpiper2005}). In consistency with the aforementioned decoupling between the excitation and main-beam angle, the optimal cavity configuration for maximal directivity is \emph{identical} to the one derived for broadside radiation (compare Tables \ref{tab:antenna_performance} and \ref{tab:antenna_performance_oblique}). Indeed, comparison of Fig. \ref{fig:antenna_results_oblique}(e)-(h) with Fig. \ref{fig:antenna_results} indicates that the field profile inside the cavity is practically independent of ${{\theta }_{\text{out}}}$. 

\begin{table*}[htb]
\begin{threeparttable}[b]
\renewcommand{\arraystretch}{1.3}
\caption{{Radiation characteristics of cavity-excited HMS antennas radiating towards $\theta_\mathrm{out}=30^{\circ}$} (corresponding to Fig. \ref{fig:antenna_results_oblique}).}
\label{tab:antenna_performance_oblique}
\centering
\begin{tabular}{l|c c c|c c c}
\hline \hline
& \multicolumn{3}{c}{$L=10\lambda$} & \multicolumn{3}{|c}{$L=14\lambda$} 
	 \\
	 & \multicolumn{3}{c}{($d=1.61\lambda, \left|z'\right|=1.00\lambda, \xi_\mathrm{out}=\frac{5}{36}\pi$)} & \multicolumn{3}{|c}{($d=1.89\lambda, \left|z'\right|=1.33\lambda, \xi_\mathrm{out}=\frac{2}{36}\pi$)}
	 \\
\hline
& Full-wave & Semianlytical & Uniform 
& Full-wave & Semianlytical & Uniform 
\\
\hline \hline \\[-1.3em]
	\begin{tabular}{l} Main beam \end{tabular}
	 & $30.2^{\circ}$ & $30.0^{\circ}$ & $30.0^{\circ}$
	 & $30.2^{\circ}$ & $30.0^{\circ}$ & $30.0^{\circ}$
	 	 	\\	\hline 	
	 \begin{tabular}{l} HPBW \end{tabular}
	 & $6.20^{\circ}$ & $5.90^{\circ}$ & $5.86^{\circ}$
	 & $4.12^{\circ}$ & $4.20^{\circ}$ & $4.20^{\circ}$
	 	 	\\	\hline 
	 \begin{tabular}{l} Directivity (2D) [dBi] \end{tabular}
	 & $16.54$ & $17.23$ & $17.36$
	 & $18.12$ & $18.49$ & $18.82$
	 	 	 	 \\	\hline 
	 \begin{tabular}{l} Side lobe $\#1$\tnote{1} \end{tabular}
	 & $20.55^{\circ}$ & $20.49^{\circ}$ & $20.49^{\circ}$
	 & $23.32^{\circ}$ & $22.96^{\circ}$ & $23.13^{\circ}$
	 	 	 	 \\	\hline 
	 \begin{tabular}{l} Side-Lobe $\#1$ Level [dB]\tnote{1} \end{tabular}
	 & $-13.7$ & $-11.4$ & $-13.5$
	 & $-9.0$ & $-9.7$ & $-13.5$
	 			\\	\hline 
	 \begin{tabular}{l} Side lobe $\#2$\tnote{1} \end{tabular}
	 & $41.04^{\circ}$ & $40.54^{\circ}$ & $40.54^{\circ}$
	 & $37.42^{\circ}$ & $37.59^{\circ}$ & $37.38^{\circ}$
	 	 	 	 \\	\hline 
	 \begin{tabular}{l} Side-Lobe $\#2$ Level [dB]\tnote{1} \end{tabular}
	 & $-12.0$ & $-12.3$ & $-13.5$
	 & $-10.2$ & $-10.4$ & $-13.5$
	 	 	 	 \\ 	
\hline \hline
\end{tabular}
\begin{tablenotes}
\item [1] Side lobes $\#1$ and $\#1$ refer, respectively, to the first side lobes at angles lower and higher than the main beam angle.
\end{tablenotes}
\end{threeparttable}
\end{table*}

To match better the range of required surface impedance values [Fig. \ref{fig:antenna_results_oblique}(a)-(b)] to the one achievable by our spider unit-cells (Appendix \ref{sec:spider_cells_modelling}), we introduce a constant phase-shift ${{\xi }_{\text{out}}}$ to the aperture fields \cite{Epstein2014, Selvanayagam2014}. This adds a constant ${{\xi }_{\text{out}}}/2$ to the argument of the cotangent surface impedance modulation [Eq. \eqref{equ:HMS_design}], which varies the distribution of required surface impedance values, but does not affect the radiation pattern. Simulation methodology follows Appendix \ref{sec:antenna_simulations}, with the difference that the PMC in the $\widehat{xz}$ plane cannot be used, as the HMS is not symmetric in this case [Fig. \ref{fig:antenna_results_oblique}(a)-(b)]. This doubles the volume of the simulation domain, resulting in insufficient convergence of the computations for $L=25\lambda $ devices. Hence, only results for $L=10\lambda $ and $L=14\lambda $ are presented.

\textcolor{black}{Table \ref{tab:antenna_performance_oblique} and Fig. \ref{fig:antenna_results_oblique} indicate that the excellent agreement between the semianalytical predictions and full-wave simulations is retained for devices radiating at oblique angles. For the $L=10\lambda$ and $L=14\lambda$ antennas, respectively, very high aperture efficiencies of $\eta_\mathrm{apr}=83\%$ and $\eta_\mathrm{apr}=85\%$ are recorded. The HPBWs are within $94\%$ and $101\%$ of the optimum corresponding to uniform illumination, comparable to the characteristics achieved for broadside-radiating cavity-excited HMS antennas.}

\section{Cavity-excited HMS antennas with reduced side-lobe level}
\label{sec:reduced_SLL}
As denoted in Subsection \ref{subsec:cavity_design}, for an optimized cavity thickness $d$, i.e. one that maximizes coupling to the highest-order mode, one may utilize the source position $z'$ as an additional degree of freedom to optimize the antenna radiation characteristics to achieve desirable performance. In Sections \ref{sec:theory} and \ref{sec:results}, we utilized this degree of freedom to suppress the coupling to the lowest-order mode, in order to facilitate the highest possible directivity. This, in consistency with array theory \cite{Balanis2008_Chap11_Arrays}, comes at the expense of side-lobe level.
To demonstrate the possibility to utilize the efficient semianalytical formulation to devise a cavity excitation which reduces the side-lobe level, we sweep the source position $z'$ for the optimal $d=1.61\lambda $ corresponding to an aperture length of $L=10\lambda $, and evaluate the directivity and side-lobe level associated with each $z'$. As we only require examination of the radiation pattern properties, it is sufficient to compute the radiated fields, which can be achieved analytically by asymptotic evaluation of Eq. \eqref{equ:transverse_fields_spectral_domain} in conjunction with Eq. \eqref{equ:cavity_source_spectrum} (See, e.g. Appendix C of \cite{Epstein2014}).

\begin{figure}[htb]
\centering
\includegraphics[width=8cm]{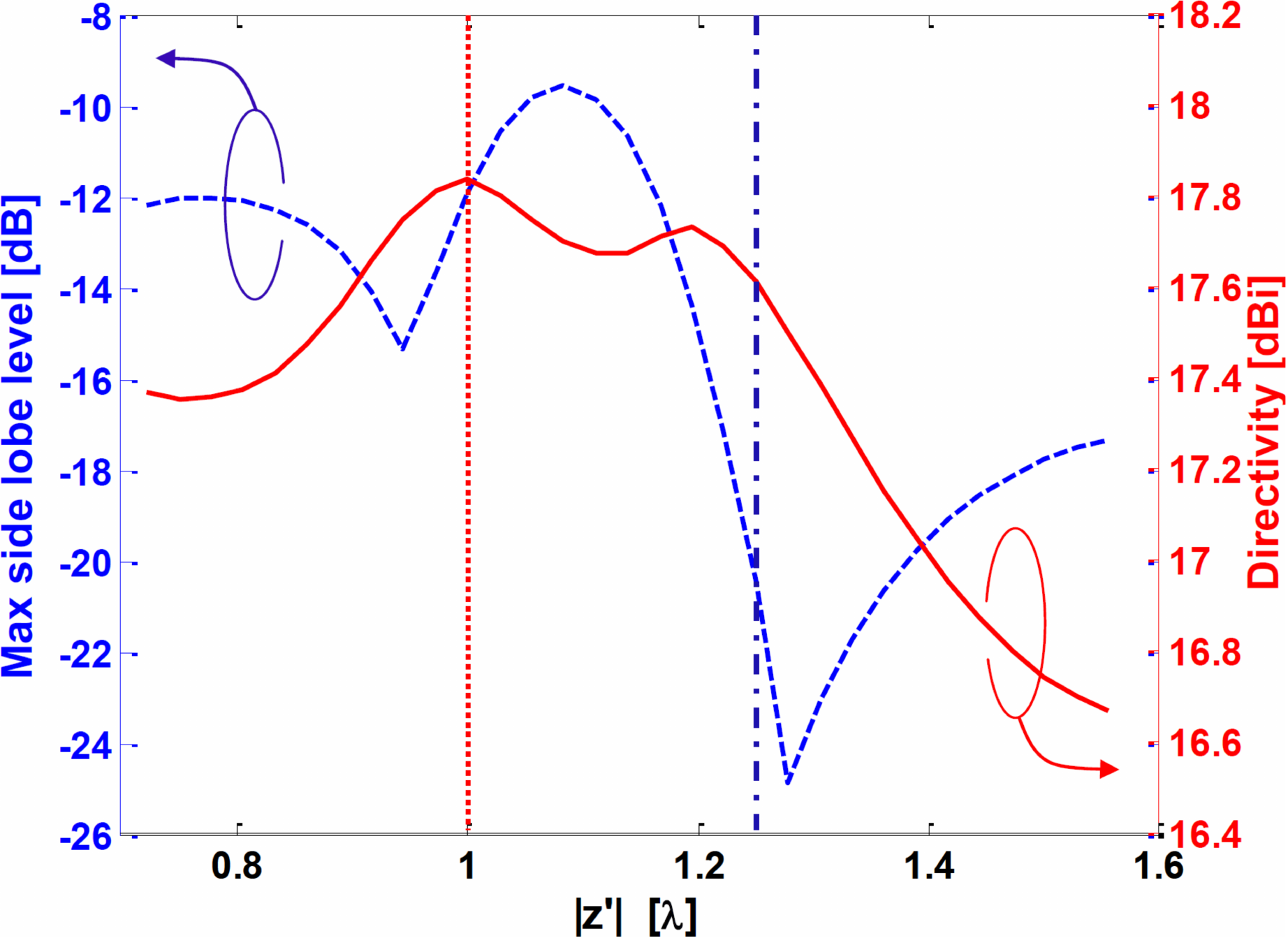}
\caption{Utilizing the source position as an additional degree of freedom for achieving desirable radiation characteristics. Directivity (red solid line) and maximal side-lobe level (dashed blue line) of a cavity-excited HMS antenna with $L=10\lambda $ and $d=1.61\lambda $ are presented as a function of source position $\left| z' \right|$. The source position maximizing directivity ($\left| z' \right|=1\lambda $, corresponding to Fig. \ref{fig:antenna_results}) is denoted by a red dotted line, and the source position for reduced side-lobe level is denoted by a blue dash-dotted line ($\left| z' \right|=1.25\lambda $, corresponding to Fig. \ref{fig:antenna_results_reduced_SLL}).
}
\label{fig:source_position_sweep}
\end{figure}

\begin{figure}[htb]
\centering
\includegraphics[width=4cm]{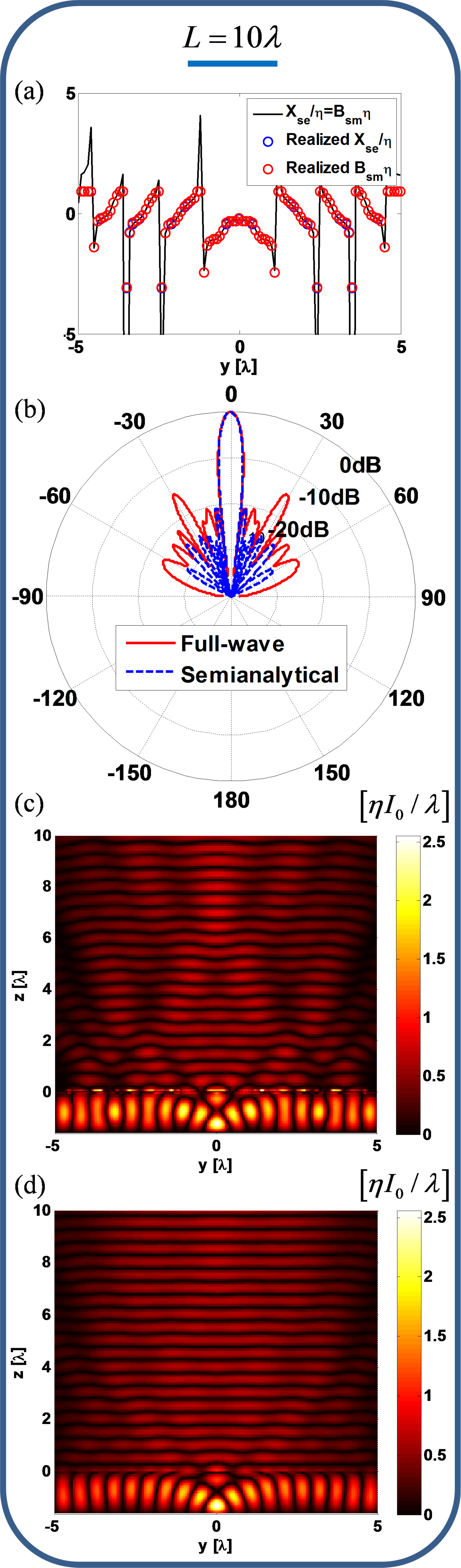}
\caption{{Performance of a cavity-excited HMS antenna with reduced side-lobe level ($L=10\lambda$).} ({a}) HMS design specifications $X_{se}\left(y\right)/\eta=B_{sm}\left(y\right)\eta$ (black solid line) derived from Eq. \eqref{equ:HMS_design}, and the realized electric surface reactance (blue circles) and magnetic surface susceptance (red circles) using the spider unit cells. ({b}) Radiation patterns produced by semianalytical formalism (blue dashed line) and full-wave simulations (red solid line). ({c}) Field distribution $\left|\Re\left\lbrace E_x\left(y,z\right)\right\rbrace\right|$ produced by full-wave simulations. ({d}) Semianalytical prediction of $\left|\Re\left\lbrace E_x\left(y,z\right)\right\rbrace\right|$ \cite{Epstein2014}.
}
\label{fig:antenna_results_reduced_SLL}
\end{figure}

The results of the parametric sweep are presented in Fig. \ref{fig:source_position_sweep}, with indication of the maximal directivity configuration ($\left| z' \right|=1\lambda $) and a configuration with reduced side-lobe level ($\left| z' \right|=1.25\lambda $). The design specifications and radiation characteristics of the cavity-excited HMS antenna corresponding to the latter are presented in Fig. \ref{fig:antenna_results_reduced_SLL} and Table \ref{tab:antenna_performance_reduced_SLL}. As expected from array theory, the reduction of side-lobe level (by about $10\mathrm{dB}$ according to full-wave simulations) results in a broadening of the main beam and a reduction of the overall aperture efficiency (although still quite high value of ${{\eta }_{apr}}=81\%$ is recorded by full-wave simulations). This demonstrates the versatility of our approach, with which a range of radiation pattern properties can be achieved.

\begin{table}[htb]
\begin{threeparttable}[b]
\renewcommand{\arraystretch}{1.3}
\caption{{Radiation characteristics of a cavity-excited HMS antenna with reduced side-lobe level} (corresponding to Fig. \ref{fig:antenna_results_reduced_SLL}).}
\label{tab:antenna_performance_reduced_SLL}
\centering
\begin{tabular}{l|c c c}
\hline \hline
& \multicolumn{3}{c}{$L=10\lambda$ ($d=1.61\lambda, \left|z'\right|=1.25\lambda$)} 
	 \\	 
\hline
& Full-wave & Semianlytical & Uniform  
\\
\hline \hline \\[-1.3em]	 	
	 \begin{tabular}{l} HPBW \end{tabular}
	 & $6.13^{\circ}$ & $5.89^{\circ}$ & $5.08^{\circ}$
	 	 	\\	\hline 
	 \begin{tabular}{l} Directivity (2D) [dBi] \end{tabular}
	 & $17.04$ & $17.62$ & $17.98$
	 	 	 	 \\	\hline 
	 \begin{tabular}{l} First Side lobe \end{tabular}
	 & $9.62^{\circ}$ & $9.21^{\circ}$ & $8.2^{\circ}$
	 	 	 	 \\	\hline 
	 \begin{tabular}{l} Side-Lobe Level [dB] \end{tabular}
	 & $-20.7$ & $-20.5$ & $-13.5$	 			
	 	 	 	 \\ 	
\hline \hline
\end{tabular}
\end{threeparttable}
\end{table}

\section{Spider unit-cell modelling}
\label{sec:spider_cells_modelling}
The spider unit cells depicted in Fig. \ref{fig:unit_cell} were defined in ANSYS Electromagnetic Suite 15.0 (HFSS 2014) with two 25mil-thick ($\approx0.64\mathrm{mm}$) Rogers RT/duroid 6010LM laminates (green boxes in Fig. \ref{fig:unit_cell}) bonded by 2mil-thick ($\approx51\mathrm{\mu m}$) Rogers 2929 bondply (white box in Fig. \ref{fig:unit_cell}). The electromagnetic properties of these products at $20\mathrm{GHz}$, e.g. permittivity tensor and dielectric loss tangent, as were provided to us by Rogers Corporation, have been inserted to the model. Specifically, a uniaxial permittivity tensor with  $\epsilon_{xx}=\epsilon_{yy}=13.3\epsilon_0, \epsilon_{zz}=10.81\epsilon_0$ and loss tangent of $\tan\delta=0.0023$ were considered for Rogers RT/duroid 6010LM laminates, while an isotropic permittivity of $\epsilon=2.94\epsilon_0$ and loss tangent $\tan\delta=0.003$ were considered for Rogers 2929 bondply. The copper traces corresponded to $\sfrac{1}{2}$ oz. cladding, featuring a thickness of $18\mathrm{\mu m}$; the standard value of $\sigma=58\times10^6\mathrm{S/m}$ bulk conductivity was used in the model. To comply with standard PCB manufacturing processes, all copper traces were 3mil ($\approx76\mathrm{\mu m}$) wide, and a minimal distance of 3mil was kept between adjacent traces (within the cell or between adjacent cells). This implies that the fixed gaps between the capacitor traces (along the $x$ axis) of the electric dipole in the middle layer, as well as between the two arms (along the $y$ axis) of the magnetic dipole in the top and bottom layer (Fig. \ref{fig:unit_cell}), were fixed to a value of $D_g=3\mathrm{mil}$ ($\approx76\mathrm{\mu m}$); the distance from the arm edge to the edge of the unit cell was fixed to $D_g/2=1.5\mathrm{mil}$ ($\approx38\mathrm{\mu m}$).

Unit cells with different values of magnetic dipole arm length $L_m$ and electric dipole capacitor width $W_e$ were simulated using periodic boundary conditions; HFSS Floquet ports were placed at $z=\pm\lambda$ and used to characterize the scattering of a normally-incident plane wave off the periodic structure (the interface between the bondply and the bottom laminate was defined as the $z=0$ plane). For each combination of $L_m$ and $W_e$, the corresponding magnetic surface susceptance $B_{sm}$ and electric surface reactance $X_{se}$ were extracted from the simulated impedance matrix of this two-port configuration, following the derivation in \cite{Selvanayagam2013_2}. 

\begin{figure}[htb]
\centering
\includegraphics[width=8cm]{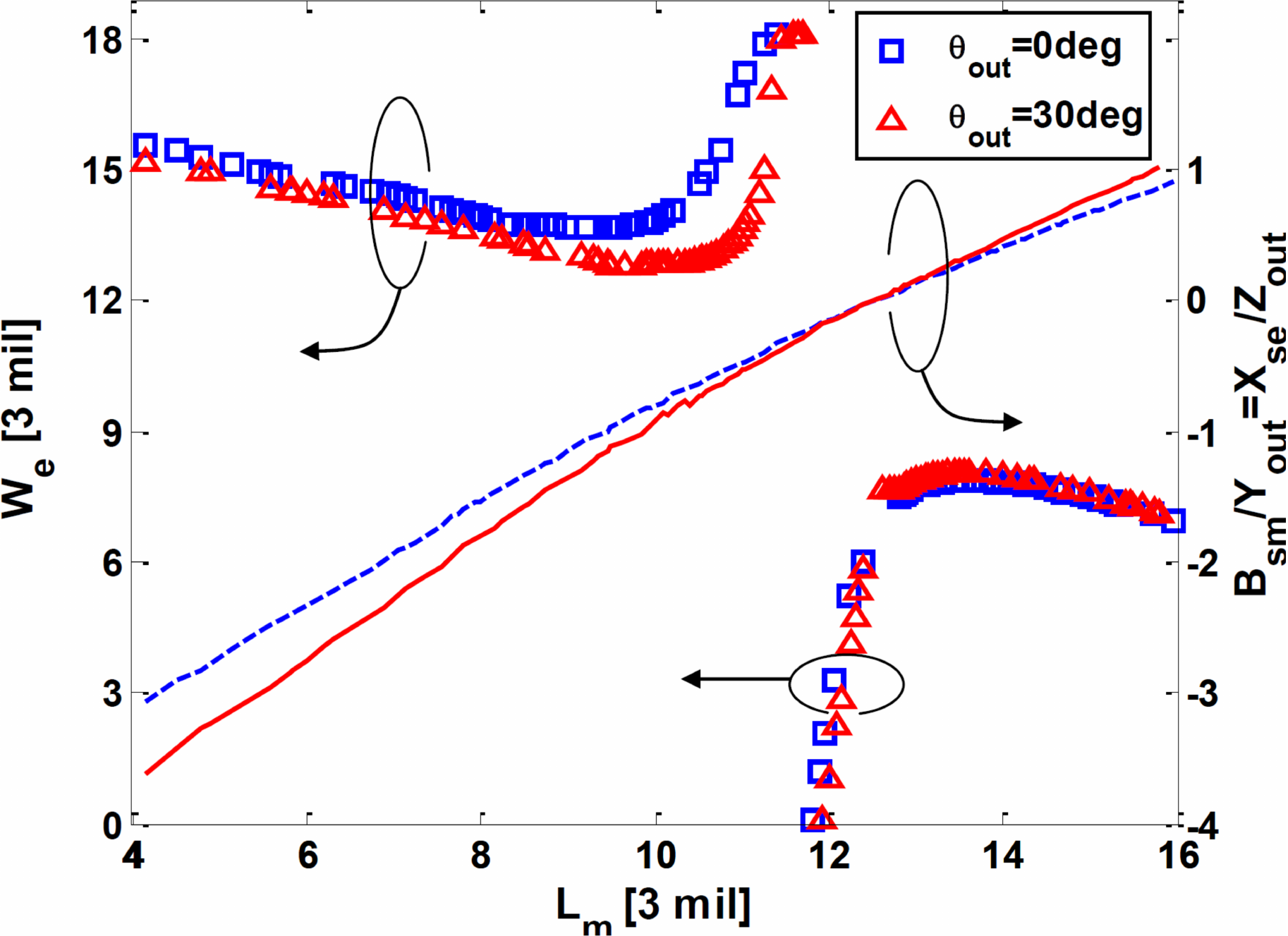}
\caption{Graphic representation of spider unit-cell lookup tables corresponding to HMSs radiating towards ${{\theta }_{\text{out}}}=0{}^\circ $ and ${{\theta }_{\text{out}}}=30{}^\circ $. Capacitor width values ${{W}_{e}}$ required for achieving balanced electric and magnetic responses ${{B}_{sm}}/{{Y}_{\text{out}}}={{X}_{se}}/{{Z}_{\text{out}}}$ are presented as a function of the magnetic dipole arm length ${{L}_{m}}$ for ${{\theta }_{\text{out}}}=0{}^\circ $ (blue open squares) and ${{\theta }_{\text{out}}}=30{}^\circ $ (red open triangles) radiators, as obtained by finite-element simulations. The corresponding ${{B}_{sm}}/{{Y}_{\text{out}}}={{X}_{se}}/{{Z}_{\text{out}}}$ values are denoted using a blue dashed line for ${{\theta }_{\text{out}}}=0{}^\circ $ (${{Z}_{\text{out}}}=1/{{Y}_{\text{out}}}=\eta $) and using a red solid line for ${{\theta }_{\text{out}}}=30{}^\circ $ (${{Z}_{\text{out}}}=1/{{Y}_{\text{out}}}=\eta /\cos 30{}^\circ $). 
}
\label{fig:spider_cell_lookup_tables}
\end{figure}

The magnetic response $B_{sm}$ was found to be proportional to the magnetic dipole arm length $L_m$, with almost no dependency in $W_e$ \cite{Wong2014}. Thus, to create an adequate lookup table for implementing \textcolor{black}{HMSs radiating towards $\theta_\mathrm{out}$}, we varied $L_m$ by constant increments, and for a given $L_m$, plotted \textcolor{black}{$B_{sm}/Y_\mathrm{out}$ and $X_{se}/Z_\mathrm{out}$} as a function of $W_e$. The value of $W_e$ for which the two curves intersected corresponded to a balanced-impedance point ($Z_{se}/Z_\mathrm{out}=Y_{sm}/Y_\mathrm{out}$), where the unit cell acts as a Huygens source, and thus suitable for implementing our metasurface. A lookup table composed of 
$\left(B_{sm},X_{se}\right)$ pairs and the corresponding unit cell geometries $\left(L_m,W_e\right)$ was constructed, and refined through interpolation. The interpolated unit cell geometries were eventually simulated again, to verify the interpolation accuracy and finalize the lookup table entries, as presented in Fig. \ref{fig:spider_cell_lookup_tables}. Finally, for a given HMS with prescribed surface impedance modulation $\left(B_{sm}\left(y\right),X_{se}\left(y\right)\right)$, a corresponding structure could be defined in HFSS using the unit cells $\left(L_m\left(y\right),W_e\left(y\right)\right)$ found via the lookup table in terms of least-squares-error.

\section{Antenna full-wave simulations}
\label{sec:antenna_simulations}
To verify our semianalytical design via full-wave simulations, each of the cavity-excited HMS antennas designed in this paper was defined in HFSS using a single strip of unit cells implementing the metasurface, occupying the region $\left|x\right|\leq \lambda/20$, $\left|y\right|\leq L/2$ ($L$ being the aperture length of the antenna), and $-0.64\mathrm{mm}\leq z \leq 0.69\mathrm{mm}$ (in correspondence to the laminate and bondply thicknesses). The simulation domain included $\left|x\right|\leq \lambda/20$, $\left|y\right|\leq L/2+2.5\lambda$, and $-d\leq z \leq 10\lambda$ ($d$ being the cavity thickness), where PEC boundary conditions were applied to the $x=\pm\lambda/20$ planes to form the equivalence of a 2D scenario. PEC boundary conditions were also applied to the $z=-d$ plane, and to two $18\mu m$-thick side-walls at $y=\pm L/2$, forming the cavity. The line-source excitation was modelled by a $\lambda/20$-wide $1\mathrm{A}$ current sheet at $z=z'$, with the current aligned with the $x$ axis. Radiation boundary conditions were applied to the rest of the simulation space boundaries, namely $z=10\lambda$, and $y=\pm\left(L/2+2.5\lambda\right)$, allowing proper numerical evaluation of the fields surrounding the antenna.

To reduce the computational effort required to solve this configuration, we utilized the symmetries of our TE scenario. Specifically, we placed a perfect-magnetic-conductor (PMC) symmetry boundary conditions at the $\widehat{xz}$ plane, and a PEC symmetry boundary conditions at the $\widehat{yz}$ plane (the PMC symmetry boundary conditions is only applicable for broadside radiators, \textit{cf.} Appendix \ref{sec:oblique_angle}). We also noticed that adding a thin layer ($1\mathrm{mil}\approx25\mathrm{\mu m}$) of copper between the electric dipole edges and the PEC parallel-plates at $x=\pm\lambda/20$ enhanced the convergence of the simulation results. With that minor modification, all of the simulated antennas converged within less than 40 iterations (maximum refinement of $10\%$ per pass), where the stop conditions was 3 consecutive iterations in which $\Delta\mathrm{Energy}<0.03$.

\section{Design procedure assumptions}
\label{sec:design_assumptions}
Several assumptions made during the derivation of the HMS design formulas contribute to discrepancies between predicted and actual performance of the presented antennas. First, the predicted fields are derived assuming the HMS is capable of implementing continuous surface impedance boundary conditions, with unbound surface impedance values; nevertheless, the physical implementation requires discretization of the continuous modulation into unit-cells, and the range of achievable surface impedance values is limited (See Fig. \ref{fig:spider_cell_lookup_tables}). Second, the HMS is assumed to be passive and lossless, however realistic conductors and dielectrics, used for the implementation of the devices in ANSYS HFSS, include unavoidable losses. Third, to facilitate the plane-wave-like relation between the transmitted fields on the aperture [Eq. \eqref{equ:envelope_function}], while still guaranteeing they obey Maxwell's equations, we have used the approximation
\begin{equation}
\left| {{\mathsf{\mathcal{F}}}^{-1}}\left\{ \tfrac{1}{2\beta }T\left( {{k}_{t}} \right) \right\} \right|\approx \left| {{\mathsf{\mathcal{F}}}^{-1}}\left\{ \tfrac{1}{2\beta }T\left( {{k}_{t}} \right)\left[ 1\pm \Gamma \left( {{k}_{t}} \right) \right] \right\} \right|
\label{equ:SVE_1}
\end{equation}
which is satisfied when	
\begin{equation}
\mathsf{\mathcal{E}}\left( y \right)\triangleq \left| \frac{{{\mathsf{\mathcal{F}}}^{-1}}\left\{ \tfrac{1}{2\beta }T\left( {{k}_{t}} \right)\Gamma \left( {{k}_{t}} \right) \right\}}{{{\mathsf{\mathcal{F}}}^{-1}}\left\{ \tfrac{1}{2\beta }T\left( {{k}_{t}} \right) \right\}} \right|\ll 1,
\label{equ:SVE_2}
\end{equation}
which is a refinement of the slowly-varying envelope (SVE) approximation utilized in \cite{Epstein2014}. This approximation is self-consistent with our design scheme, as when the transmitted fields are directive towards   ${{\theta }_{\text{out}}}$ as desirable, the dominant components of the transmission spectrum $T\left( {{k}_{t}} \right)$ are in the vicinity of ${{k}_{t}}=k\sin {{\theta }_{\text{out}}}$, where the reflection coefficient $\Gamma \left( {{k}_{t}} \right)$ completely vanish (the numerator of Eq. \eqref{equ:SVE_2} vanishes).

\begin{figure*}[htb]
\centering
\includegraphics[width=11cm]{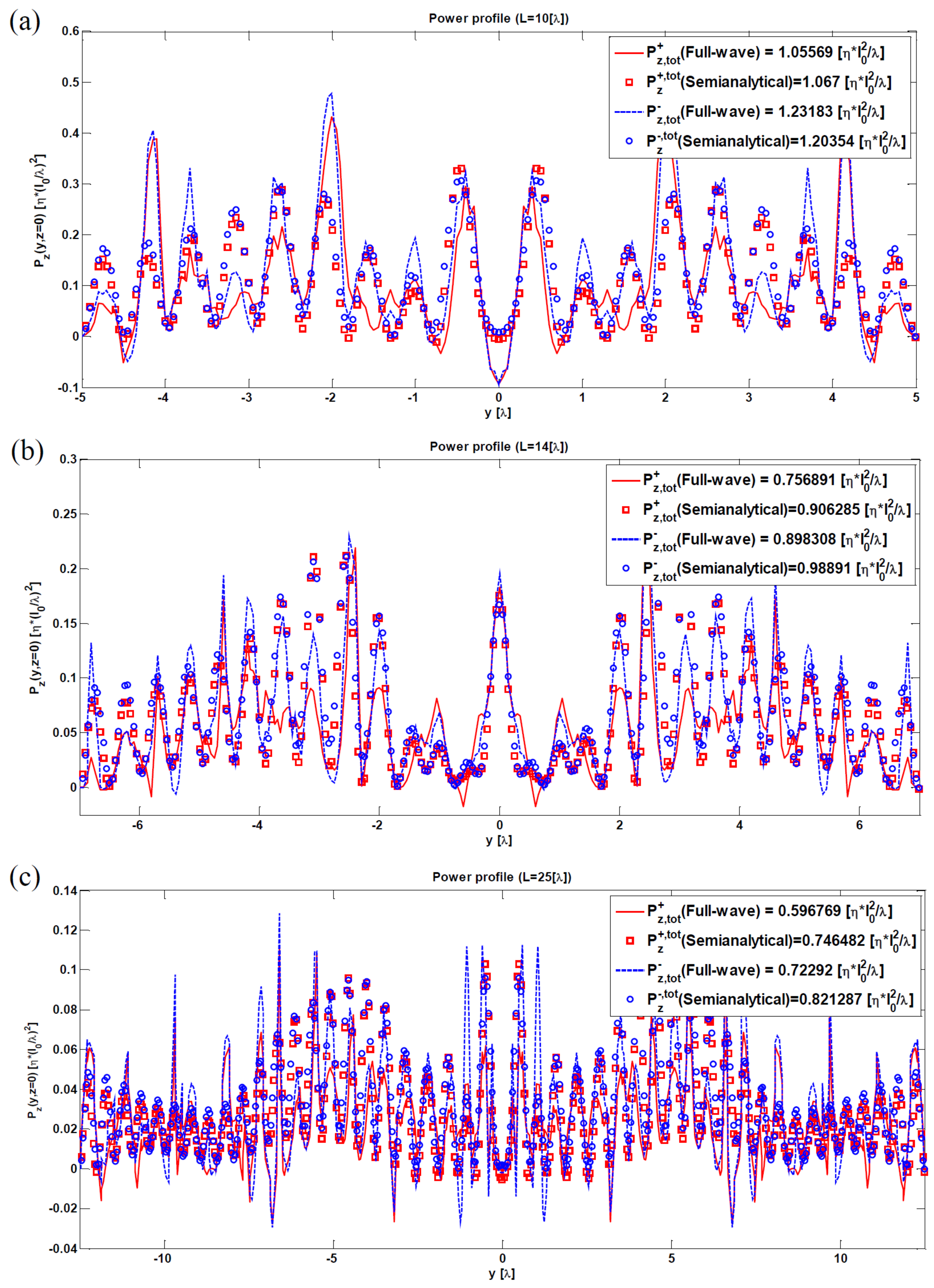}
\caption{Power profiles on the cavity-excited HMS antenna apertures. The real part of the $z$-component of the Poynting vector $P_{z}^{\pm }=\Re \left\{ {{S}_{z}}\left( y,z\to {{0}^{\pm }} \right) \right\}$ above and below and the metasurface is presented for the antennas reported in Table I and Fig. 5, with aperture lengths (a) $L=10\lambda $ (b) $L=14\lambda $, and (c) $L=25\lambda $. Full-wave simulation results of $P_{z}^{+}$ (red solid line) and $P_{z}^{-}$ (blue dashed line) were evaluated at $z=\lambda /10$ and $z=-\lambda /10$, respectively. Semianalytical predictions of  $P_{z}^{+}$ (red open squares) and $P_{z}^{-}$ (blue open circles) are presented as well. The total power $P_{z,\text{tot}}^{\pm }$, calculated by integrating the power profile along the aperture is indicated in the legend.
}
\label{fig:power_profiles}
\end{figure*}

Interestingly, the impacts of these three assumptions can be assessed by reviewing the predicted and simulated power flow across the metasurface. The $z$-directed power profiles below (blue) and above (red) the HMS as predicted by the semaianalytical formalism (open circles and squares, respectively) and as extracted from full-wave simulations (dashed and solid lines, respectively) are presented in Fig. \ref{fig:power_profiles}, for the three antennas reported in Section \ref{sec:results}. The fact that the general trend and quantitative data of the semianalytical and simulated results compare well (note that the profiles are plotted using a common $\eta {{\left( {{I}_{0}}/\lambda  \right)}^{2}}$ unit scale), indicates that the first assumption is valid (this is also supported by \cite{Epstein2014_2}). The semianalytical predictions made based on a homogenized continuous surface impedance boundary conditions mostly agree with the simulation data recorded  below and above the metasurface, where effective medium theory predicts discretization effects to be negligible \cite{Tretyakov2003,Epstein2014}.

Violations of the second assumption, regarding the lossless nature of the HMS, would manifest themselves as differences between the simulated power profile below and above the metasurface, which must originate in dissipation in the unit-cell conductors and dielectrics. On the other hand, violations of the third assumption, related to the SVE approximation, would manifest themselves as differences in the semianalytically predicted power profile below and above the metasurface, as they correspond to violations of local power conservation \cite{Epstein2014}.

\textcolor{black}{Although} local deviations from these two assumptions are found to be rather small, they contribute to a non-negligible reduction of the total power flow across the metasurface (integrated over the aperture length). The values denoted in the legends of Fig. \ref{fig:power_profiles} indicate that according to full-wave simulations about $15\%$ of the power \textcolor{black}{available} below the HMS is dissipated in the lossy conductors and dielectrics, while the semianalytically predictions reveal about $10\%$ discrepancy between the power below and above the metasurface. 

While these relative deviations can be considered small albeit non-negligible, it seems that they actually balance each other. The theoretical derivation assumes and prescribes a lossless HMS, but the minor violations of the SVE approximation contribute to predicted (Maxwellian) fields which must be supported by small losses. On the other hand, the implemented HMS does include realistically unavoidable losses, which turn out to dissipate a comparable amount of power. We hypothesize that this balance allows overcoming the minor deviations from the theoretical assumptions, facilitating the very good agreement between predicted and simulated results reported herein.

\end{document}